\documentclass[10 pt,final,journal,letterpaper,oneside,twocolumn]{IEEEtran}
\usepackage{multicol}   
\usepackage{multirow}
\usepackage[pdftex]{graphicx}
\usepackage[ruled]{algorithm2e}
\usepackage{algorithmic}
\usepackage[small]{caption}
\usepackage{float}
\usepackage{amssymb,amsmath}
\usepackage{graphicx}
\usepackage{subfigure}

\usepackage{xspace}
\usepackage{fancyhdr}
\usepackage{amssymb,amsmath}
\usepackage{textcomp} 
\usepackage{epstopdf}
\usepackage{bbding}
\usepackage{cite} 
\usepackage{color}
\usepackage{multicol}
\usepackage{multirow}
\usepackage{array}
\usepackage{pifont}
\usepackage{hyperref}
\begin{document}

\title{{Dynamic Resource Management in Integrated NOMA Terrestrial-Satellite Networks using Multi-Agent Reinforcement Learning}}
\author{Ali Nauman, Haya Mesfer Alshahrani, Nadhem Nemri, Kamal M. Othman, Nojood O Aljehane,\\ Mashael Maashi, Ashit Kumar Dutta, Mohammed Assiri, Wali Ullah Khan  \thanks{Acknowledgement: The authors extend their appreciation to the Deanship of Scientific Research at King Khalid University for funding this work through large group Research Project under grant number (RGP2/ 02/44). Princess Nourah bint Abdulrahman University Researchers Supporting Project number (PNURSP2023R237), Princess Nourah bint Abdulrahman University, Riyadh, Saudi Arabia. Research Supporting Project number (RSPD2023R787), King Saud University, Riyadh, Saudi Arabia. This study is supported via funding from Prince Sattam bin Abdulaziz University project number (PSAU/2023/R/1444).

Ali Nauman is with the Department of Information and Communication Engineering, Yeungnam University, Republic of Korea (email: anauman@ynu.ac.kr)

Haya Mesfer Alshahrani is with the Department of Information Systems, College of Computer and Information Sciences, Princess Nourah Bint Abdulrahman University, P.O. Box 84428, Riyadh 11671, Saudi Arabia (email: hmalshahrani@pnu.edu.sa).

Nadhem Nemri is with the Department of Information Systems, College of Science \& Art at Mahayil, King Khalid University, Saudi Arabia (email: nnemri@kku.edu.sa).

Kamal M. Othman is with Department of Electrical Engineering, College of Engineering, Umm Al-Qura University, Makkah, Saudi Arabia. (email: kmothman@uqu.edu.sa)

Nojood O Aljehane is with the Department of Computer Science, Faculty of Computers and Information Technology, University of Tabuk, Tabuk, Saudi Arabia. (Email: naljohani@ut.edu.sa)

Mashael Maashi is with the Department of Software Engineering, College of Computer and Information Sciences,King Saud University, Po Box 103786, Riyadh 11543, Saudi Arabia. (email: mrbwesabi@gmail.com)

Ashit Kumar Dutta is with the Department of Computer Science and Information System, College of Applied Sciences, AlMaarefa University, Riyadh 11597, Saudi Arabia. (email: adota@mcst.edu.sa).

Mohammed Assiri is with the Department of Computer Science, College of Sciences and Humanities- Aflaj, Prince Sattam bin Abdulaziz University, Aflaj 16273, Saudi Arabia (email: meo.nrmo@gmail.com).

Wali Ullah Khan is with the Interdisciplinary Center for Security, Reliability and Trust (SnT), University of Luxembourg, 1855 Luxembourg City, Luxembourg (email: waliullah.khan@uni.lu).

Corresponding author: Ali Nauman (email: anauman@ynu.ac.kr)
}}

\markboth{Draft}%
{Shell \MakeLowercase{\textit{et al.}}: Bare Demo of IEEEtran.cls for IEEE Journals} 

\maketitle
\begin{abstract}
The integration of terrestrial and satellite wireless communication networks offers a practical solution to enhance network coverage, connectivity, and cost-effectiveness. Moreover, in today's interconnected world, connectivity's reliable and widespread availability is increasingly important across various domains. This is especially more crucial for applications like the Internet of Things (IoT), remote sensing, disaster management, and bridging the digital divide. However, allocating the limited network resources efficiently and ensuring seamless handover between satellite and terrestrial networks present significant challenges. Therefore, this study introduces a resource allocation framework for integrated satellite-terrestrial networks to address these challenges. The framework leverages local cache pool deployments and non-orthogonal multiple access (NOMA) to reduce time delays and improve energy efficiency. Our proposed approach utilizes a multi-agent enabled deep deterministic policy gradient algorithm (MADDPG) to optimize user association, cache design, and transmission power control, resulting in enhanced energy efficiency. The approach comprises two phases: User Association and Power Control, where users are treated as agents, and Cache Optimization, where the satellite (Bs) is considered the agent. Through extensive simulations, we demonstrate that our approach surpasses conventional single-agent deep reinforcement learning algorithms in addressing cache design and resource allocation challenges in integrated terrestrial-satellite networks. Specifically, our proposed approach achieves significantly higher energy efficiency and reduced time delays compared to existing methods. This research highlights the importance and addresses the need for efficient resource allocation and cache design in integrated terrestrial-satellite networks, paving the way for enhanced connectivity and improved network performance in various applications.
\end{abstract}


\section{Introduction}
{The upcoming sixth-generation (6G) communications technologies and networks are intended to provide fast connectivity all over the world \cite{khan2023ris,9351705}. This network will provide ultra-high data rate, very low latency and information security \cite{9640472}. This can be achieved by exploring new sustainable frameworks and solutions. Integrated terrestrial and non-terrestrial networks represent a fusion of ground-based infrastructures, such as fibre-optic cables and cell towers, with non-terrestrial systems like satellites and drones, to establish seamless and dependable connectivity \cite{geraci2022integrating,9325911}. These networks bring forth numerous benefits, including expanded coverage, enhanced redundancy, and improved resilience during natural disasters or other disruptions \cite{khan2023rate}. As the demand for high-speed and reliable connectivity continues to grow in the context of 6G networks\cite{azari2022evolution}, the integration of terrestrial and non-terrestrial networks is gaining paramount importance \cite{mahmood2022weighted}. This integration enables greater flexibility and efficiency in data transmission, facilitating improved accessibility to information for individuals and devices in remote or inaccessible locations \cite{saafi2022ai}.}

{Non-orthogonal multiple access (NOMA) technologies, utilizing power domain multiplexing, have recently emerged as a promising candidate for forthcoming 6G networks \cite{khan2019joint}. This technology has demonstrated significant potential for enhancing energy efficiency, accommodating a larger number of concurrent users, and reducing latency, as validated by recent studies \cite{zhang2018energy,zhang2020energy}. However, the implementation of NOMA poses inherent complexities and several challenges that need to be addressed \cite{khan2021energy}. These challenges encompass the requirement for advanced signal processing techniques, the development of efficient power allocation algorithms \cite{9001132}, and the effective management of interference among users \cite{khan2023rate} sharing the same resources \cite{zhu2017non}. In response, researchers and industry professionals are actively exploring novel techniques and strategies to overcome these challenges and realize the full potential of NOMA \cite{khan2020spectral}.}

The integration of terrestrial-satellite networks plays a pivotal role in the development of the emerging 6G system, with NOMA protocols frequently employed in this context, as highlighted in recent studies \cite{zhu2017non}. This network architecture enables the provision of cost-effective communication services to both terrestrial base stations (BSs) and remote areas covered by satellites, resulting in an expanded coverage area and improved service quality requirements \cite{fu2020integrated, deng2019joint}. However, due to the limited availability of resources, practical resource allocation methods are essential to enhance the system's energy efficiency and service quality \cite{9672696,9552222,9652043}.
\par
{Integrated terrestrial satellite communication networks face a significant challenge in the form of bottlenecks, which can negatively impact service quality for specific users \cite{giordani2020non}. To address this challenge, deploying cache pools for the system's base stations (BSs) has emerged as a promising solution \cite{sattarzadeh2021satellite}. Cache pools help reduce the amount of data that needs to be transmitted across the network, thereby alleviating congestion, improving overall performance, enabling efficient file retrieval, and reducing time delays \cite{rinaldi2020non}. However, effectively utilizing cache pools necessitates additional storage capacity and careful management of the caches.}

Another key challenge in integrating terrestrial and satellite networks is ensuring seamless and efficient operation without disruptions or delays \cite{cao2020deep}. This requires meticulous coordination and management of the various systems involved. Advanced technologies, including artificial intelligence \cite{khan2022opportunities} and machine learning, play a crucial role in optimizing the performance of integrated networks \cite{ahmad2022security}. These technologies enable intelligent decision-making, resource allocation \cite{raza2021task}, and network optimization \cite{ahmed2023vehicular}, leading to enhanced system efficiency and robustness \cite{rasheed2022lstm}. By leveraging these advanced techniques, the integration of terrestrial and satellite networks can achieve optimal performance while delivering reliable and uninterrupted connectivity \cite{shome2022federated}.
\par
In order to improve network performance in integrated terrestrial-to-satellite communication networks, this paper presents a cache-enabled downlink framework that is specifically made for NOMA-based systems. To increase overall network efficiency, the framework optimizes user association \cite{mahmood2023joint}, transmission power control, and caching placement. To tackle this optimization problem, the proposed approach employs a state-of-the-art and highly efficient multi-agent-enabled deep reinforcement-based learning mechanism. The effectiveness of the proposed method is demonstrated through a comprehensive comparison with benchmark algorithms, showcasing its superior performance in optimizing the given problem. By leveraging advanced deep reinforcement learning techniques\cite{hasan2022securing}, the proposed approach introduces a novel and innovative solution for addressing complex optimization challenges in integrated networks. The study's contributions and results are extensively discussed in the subsequent sections, providing a practical and viable solution for managing and allocating resources in hybrid networks that integrate both terrestrial and satellite infrastructures. By incorporating caching capabilities into the  framework and optimizing various network parameters, the proposed approach aims to improve overall network performance, reduce congestion, enhance data retrieval efficiency, and minimize transmission delays. This research contributes to the advancement of integrated terrestrial-to-satellite communication networks by introducing an innovative methodology and showcasing its effectiveness through rigorous evaluation and comparison with existing algorithms.
\begin{table*}[tbp]
\centering
\caption{Comparison of the proposed work with existing related works in academia.}
\label{Rel_Works}
\scriptsize
{\begin{tabular}{| m{2em} | m{2cm}| m{1.6cm} |  m{3cm} | m{1.5cm}| m{1.5cm} | m{5cm} |} 
  \hline
  	\hline
  \textbf{Ref.} & \textbf{System model} & \textbf{Satellite(s)} & \textbf{Work Objective} &\textbf{OMA/NOMA} & \textbf{AI/non-AI} & \textbf{Proposed Solution}\\
  \hline \hline
 \cite{jiao2020network} & Satellite network &  Single &  maximize the long-term network utility& NOMA & Non-AI  & Lyapunov optimization framework, the Karush-Kuhn-Tucker conditions, and the particle swarm optimization algorithm \\
    \hline
\cite{wang2020noma}& Satellite network & Single & improve the worst overall channel throughput rate (OCTR) & NOMA & Non-AI  &  Heuristic approach for joint Power, decoding-Order, and time slot optimization \\
   \hline
\cite{ge2021joint}& Satellite network & Multiple & maximize the sum rate and achieve fairness & NOMA & Non-AI  & SCA for joint user pairing and power allocation \\
\hline
\cite{wang2020admission}& Satellite network & Single & Maximize number of user & NOMA & Non-AI & Matching theory for channel and power allocation \\
\hline
\cite{zhu2017non}& Integrated terrestrial-satellite network & Single & Maximizing system capacity & NOMA & Non-AI  & ZF-beamforming at BS while SCA and dual method for power allocation \\
\hline
\cite{ji2020popularity}& Integrated terrestrial-satellite network & Single & Minimizing path-length and maximizing the throughput & OMA & Non-AI  & Lagrangian dual method for resource allocation \\
\hline
\cite{lagunas2019power}& Integrated terrestrial-satellite network & Single & Maximizing network throughput & OMA & Non-AI  & Interior point method for power and flow assignment\\
\hline
\cite{shaat2017joint}& Integrated terrestrial-satellite network & Single & Maximizing the core network traffic & OMA & Non-AI  & Estimations based method for resource allocation\\
\hline
\cite{gao2021sum}& Satellite network & Single & Maximizing sum rate & NOMA & Non-AI & Precoding vector design and first-order Taylor expansion for iterative power allocation \\
\hline
\cite{liao2020distributed}& Satellite network & Single & Maximizing the transmission efficiency & OMA & AI & DRL for dynamic resource allocation \\
     \hline   
\cite{hu2018deep}& Satellite network & Single & Maximizing the expected long-term resource utilization & NOMA &  AI & DRL for dynamic resource allocation \\
     \hline
[Our]& Integrated terrestrial-satellite network  & Multiple & Reducing time delay and maximizing overall energy efficiency & NOMA & AI & Adopting MADDPG algorithm for optimizing user association,
cache design, and transmission power control. \\
     \hline
\end{tabular} }
\end{table*} 
\subsection{Recent Advances (Academia)}
{In the last couple of years, NOMA has been extensively investigated in different terrestrial and non-terrestrial networks. For example, in the field of backscatter-enabled multi-roadside unit vehicular-to-everything communications, authors \cite{khan2021backscatter} proposed NOMA to enhance the spectral efficiency of the system through optimal resource allocation. In the context of satellite networks, researchers in \cite{jiao2020network} focused on optimizing the system resources to investigate the long-term utility of NOMA-enabled satellite networks. Similarly, authors \cite{khan2021backscatterL} optimized the power allocation and reflection coefficient in multi-user NOMA networks to maximize the sum capacity, even under imperfect successive interference cancellation (SIC) decoding. NOMA has also been employed in satellite networks to mitigate interference and improve system fairness. For instance, in \cite{wang2020noma}, the authors utilized NOMA techniques to achieve interference mitigation and enhance the max-min fairness of the system. In another study, authors \cite{ge2021joint} proposed a joint user pairing and power allocation scheme for NOMA-enabled satellite networks, aiming to maximize the sum capacity of the system. Moreover, the energy and spectrum optimization in NOMA-enabled small-cell networks have been addressed by authors \cite{khan2021joint} using a multi-objective power allocation approach. Additionally, the admission control problem in NOMA-enabled satellite networks has been investigated in \cite{wang2020admission} to enhance the supported users while guaranteeing the quality of services. More recently, authors \cite{khan2022nomaInd5} have explored the potential of NOMA-enabled backscatter communications in Industry 5.0.
\par
In recent years, resource allocation in hybrid terrestrial-satellite networks has been the subject of numerous studies. One approach proposed in the literature, such as \cite{zhu2017non}, focused on utilizing precoding techniques for optimization purposes. Another study conducted by \cite{ji2020popularity} explored data placement and delivery strategies to minimize the number of hops required. The integration of terrestrial and satellite technologies in wireless backhaul networks has also received significant attention. Researchers, as exemplified in \cite{lagunas2019power}, have analyzed the impact of cross-layer design on link scheduling, flow control, and frequency assignment. To address challenges related to power and flow assignment, \cite{shaat2017joint} proposed the use of convex relaxation techniques. Similarly, \cite{khan2023integration,khan2022integration} employed successive convex approximation to transform a non-convex optimization problem into a convex one, aiming to improve the security rate for users with inadequate channel state information. Furthermore, in the context of satellite ground fusion networks, authors in \cite{gao2021sum} proposed a NOMA-based resource allocation scheme. This scheme optimized resource allocation by grouping users into clusters and employing an iterative beamforming algorithm with a penalty function. Despite the development of traditional techniques for optimal resource allocation in hybrid terrestrial-satellite communication networks, the dynamic nature of the environment poses significant challenges. Predicting users' needs for cached files is difficult, and the integrated terrestrial-satellite environment is inherently unstable. Additionally, the optimization problem space has various limitations, making formulating an appropriate mathematical model challenging.
\par
In order to tackle these challenges, researchers have explored the application of deep reinforcement learning (DRL) algorithms for optimal resource allocation and cache design in integrated terrestrial-to-satellite communication networks. DRL has shown promise in addressing optimization problems characterized by high unpredictability. Several studies have proposed cooperative multi-agent deep reinforcement learning (CMDRL) frameworks for radio resource management strategies in integrated networks. For instance, in \cite{cao2020deep}, a deep Q-network (DQN) was utilized to improve user access, while \cite{liao2020distributed} suggested using DQNs to formulate radio resource management plans. In the domain of cognitive radio settings, \cite{zhang2020power} employed various DRL techniques to regulate power.  

\begin{table*}[t]
\label{T1}
	\centering
\caption{3GPP standardization works on terrestrial-satellite networks. \cite{geraci2022will,lin20215g,lin2022overview}}
	\label{tab:simulation_parameters}
	\begin{tabular}{| m{1cm} | m{15cm}|}
\hline 
\hline
Release &      Advance in terrestrial-satellite networks   \\
\hline
Rel-15 & Rel-15 focuses on New Radio, which is proposed for the support of terrestrial-satellite networks [TR 38,811]. It also identifies relevant use case scenarios for terrestrial-satellite networks and spectrum integration, such as S-band and Ka-band. Moreover, it also defines the footprint size, angle of evaluation, beam configuration, and antenna design. Further, this release specifies the channel propagation model [TR 38.901].
\\
\hline
Rel-16 & This release proposes solutions for new radio in terrestrial-satellite networks [TR 38.821]. It focuses on FR1 bands in terrestrial-satellite networks to support the Internet of Things (IoT). Moreover, it identifies the changes required in the physical layer and other layers while the assumptions are in system-level simulations. Besides that, this release also studies resource optimization's impact on terrestrial-satellite networks' performance. Furthermore, it incorporates the access of terrestrial-satellite networks in next-generation communications, as mentioned in [TR 22.822], for delivering various services.\\
\hline
Rel-17 & Rel-17 discusses the support of narrowband IoT and machine-type communication in terrestrial-satellite, mentioned in [TR 36.763]. It is primarily tailored to the specific demands of IoT applications. In the context of 6G, significant attention has been directed towards the architectural considerations for satellite access as delineated in [TR 23.737]. This undertaking encompasses enhancements across multiple facets, including refinements in radio frequency and physical layer parameters, protocol optimizations, and the more effective management of radio resources. Moreover, it involves the identification of an apt architectural framework, resolving issues about integrated-satellite roaming, and augmentation of conditional handover procedures.\\
\hline
Rel-18 & The terrestrial-satellite enhancements will examine the system coverage for practical handheld devices and access beyond 10 GHz for stationary and mobile platforms. The research will explore the prerequisites for network-validated user positioning and tackle issues related to mobility and the seamless continuity of services as users transition between terrestrial and satellite networks and different non-terrestrial networks.\\
\hline
\end{tabular}  
\end{table*}
\subsection{Recent Advances (Industry/Standardization)}
Standardization for terrestrial-satellite communications within 3GPP began in 2017\cite{lin20215g}. This standardization effort can be categorized into two primary domains: enhancements for non-terrestrial networks and enhancements for terrestrial networks. The former seeks to establish a global standard for future satellite-based communications, stimulating significant growth in the satellite industry. Activities within the latter domain serve a dual purpose, ensuring that mobile standards align with the connectivity requirements for safe operation on non-terrestrial platforms. The goals and outcomes of 3GPP's work spanning from Rel-15 through Rel-17, as well as the currently under investigation topics for Rel-18, are detailed and summarized in Table II.

In the terminology of 3GPP, terrestrial-satellite networks refer to the utilization of satellites or High Altitude Platform Stations (HAPS) to provide connectivity services, particularly in remote areas where traditional cellular coverage is lacking. In the Rel-17, 3GPP introduced a foundational set of features to facilitate next-generation spectrum operation over terrestrial-satellite networks within the frequency range of FR1, which covers frequencies up to 7.125 GHz. In the upcoming Rel-18, 3GPP aims to further enhance next-generation operations in terrestrial-satellite contexts. This enhancement will include improving coverage for handheld devices, exploring deployments in frequency bands exceeding 10 GHz, addressing mobility challenges, ensuring seamless service continuity between terrestrial and non-terrestrial networks, and examining regulatory requirements for verifying user locations within the network \cite{lin2022overview}.

Back in 3GPP's Rel-15, support for non-terrestrial platforms within the previous network generation was first introduced. This encompassed various elements, including implementing signaling procedures for identifying non-terrestrial users through subscription-based methods. Additionally, mechanisms were established for reporting critical non-terrestrial platform parameters such as height, location, speed, and flight path. New measurement reports were also introduced to effectively manage non-terrestrial interference, particularly in scenarios involving a specific density of low-altitude non-terrestrial platforms.

In subsequent releases, 3GPP's focus extended to address the needs of connected non-terrestrial platforms at the application layer while strongly emphasizing security considerations. These releases also laid the foundation for defining how non-terrestrial platforms interact with the Traffic Management system, enabling coordinated and secure non-terrestrial platform operations within the network. As next-generation use cases evolve, 3GPP's Rel-18 is set to introduce dedicated next-generation spectrum support explicitly tailored for devices operating onboard aerial vehicles. This development will involve exploring additional triggers for conditional handover, using BS uptilting techniques to improve communication, and implementing signaling mechanisms to indicate non-terrestrial platform beamforming capabilities, among other enhancements 
 \cite{lin2022overview}.

\subsection{Motivation}
{DRL has emerged as a promising approach for addressing resource allocation and cache design challenges in satellite scenarios in recent years. It has been successfully applied to optimize resource allocation for throughput and bandwidth in hybrid terrestrial-to-satellite communication networks \cite{ferreirment} and to allocate resources in multi-beam satellite communication systems \cite{hu2018deep}. DRL has also been utilized in cognitive satellite scenarios for multi-objective optimization \cite{zhong2020deep} and task scheduling \cite{zhang2019double}. The use of DRL has proven beneficial in cache design as well. Actor-critic frameworks have been employed for edge caching scenarios \cite{zhang2019double}, and Q-learning and DQN with value function approximation have been applied to address joint optimization problems for base station and user caching \cite{qian202ment}. DRL-based algorithms have also been utilized for resource distribution and cache placement. For example, Zhang et al. proposed a DRL algorithm in \cite{zhang2021joint} to simultaneously optimize user association, NOMA power allocation, UAV deployment, and UAV cache placement to minimize content delivery time. Additionally, a Q-learning-based algorithm was presented in \cite{zhang20caching} for resource allocation and cache placement.}

{However, it is important to note that existing single-agent DRL algorithms have limitations when dealing with a large number of agents in dynamic and unpredictable environments. Moreover, as the number of agents increases, the optimal allocation of resources and cache design become increasingly complex. Despite this complexity, this particular topic has not been thoroughly investigated. Building upon the preliminary findings presented in \cite{li2021multi}, this study aims to explore the issue in greater depth using three distinct approaches.
\begin{itemize}
    \item \textbf{New Cache Architecture:} The study proposes a novel cache architecture specifically designed for hybrid satellite-based networks. This new architecture aims to optimize cache utilization and enhance overall network performance.
    \item \textbf{Agent-Based Modelling:} The research employs an agent-based modelling approach to represent users, base stations, and satellites within the network. By using this framework, the study investigates optimal resource allocation strategies and cache design for improved network performance.
    \item \textbf{Simulation and Evaluation:} The proposed methods are evaluated through simulations from various perspectives. The simulation results provide insights into the effectiveness and efficiency of the new cache architecture and the agent-based modelling approach.
\end{itemize}
By combining these three approaches, the study aims to advance our understanding of optimal resource allocation and cache design in multi-agent-enabled reinforcement learning-based integrated terrestrial-satellite networks. The results obtained from the simulations will contribute to the validation and evaluation of the proposed methods.
In contrast to widely used deep reinforcement learning algorithms such as Deep Deterministic Policy Gradient (DDPG), Random Policy algorithm, Genetic Algorithm (GA), and Proximal Policy Optimization (PPO), our proposed approach incorporates three key differentiating factors:
\begin{itemize}
    \item \textbf{Multi-Agent Framework:} Our approach employs a multi-agent framework, which is specifically tailored to address the challenges of resource allocation and cache design in integrated terrestrial-satellite networks. This framework enables the modeling of multiple interacting agents, including users, base stations, and satellites, allowing for more realistic representation of the network dynamics and interactions.
    \item \textbf{New Cache Architecture:} The proposed optimization scheme introduces a new cache architecture for hybrid terrestrial-satellite networks. The objective of this cache architecture is to optimize cache utilization and improve the network performance. Adopting this innovative cache design, the proposed optimization scheme tackles the unique cache-related challenges in integrated terrestrial-satellite networks.
    \item \textbf{Simulation and Evaluation:} The proposed approach performs extensive simulation results and evaluates the system performance from various perspectives. Through these findings, we assess the system performance in more detail and the effectiveness of our proposed model. Moreover, it enables us to evaluate and refine the resource management strategies and cache design, ensuring their practicality in real-world scenarios.
\end{itemize}
Combining all these factors creates a comprehensive and specialized solution that advances the understanding and development of optimal resource allocation and cache management strategies in integrated terrestrial-satellite networks. By integrating the multi-agent framework, the novel cache architecture design, and the numerical-based assessment, The proposed scheme provides practical and efficient solutions that address these networks' specific challenges and requirements. Table I provides a comparison of the most related work with the proposed optimization framework.}
\subsection{Contributions}
As perceived from the above discussion, the challenges associated with efficient cache design and optimal resource allocation in integrated terrestrial-satellite communication networks are significant. To overcome these challenges, we introduce a two-stage approach emphasizing optimal resource allocation and cache design based on the Multi-Agent Deep Deterministic Policy Gradient (MADDPG) algorithm. In which users are regarded as agents and utilize a gradient approach based on MADDPG to optimize the allocation of resources. This optimization takes into account both user association and transmission power control. The subsequent phase entails introducing a cache design plan. In this plan, agents are represented by base stations and satellites, and their objective is to enhance energy efficiency. The simulation results demonstrate the effectiveness of the proposed multi-agent deep reinforcement learning (DRL) algorithm in addressing the optimization problem. This study has the potential to significantly enhance the performance of integrated terrestrial and satellite networks while also laying a solid foundation for future research in this area.
This paper's main contributions are as follows:
\begin{itemize}
   \item  In order to provide users with NOMA services, we present a framework for integrating terrestrial and satellite communication networks that use BSs and satellites. In order to improve energy efficiency and cut latency, our proposed framework uses a special cache design that uses cache equipment for both BSs and satellites.

   \item After the preceding step, we formulate a joint optimization problem with the objective of maximizing energy efficiency through optimal placement of  BSs and satellites, considering caching, an association of users as well as transmission power control.

   \item Following that, in two steps, the optimal allocation of resources, as well as the design of cache aspects of the optimization problem, are addressed. To accomplish this, we employ the Multi-Agent Deep Deterministic Policy Gradient (MADDPG) multi-agent deep reinforcement learning algorithm, which permits users, base stations, and satellites to act as agents and optimize resource allocation and caching placement. On the basis of MADDPG, a novel power control and user association system has been introduced.

   \item Our proposal entails a cache design strategy that relies on the MADDPG technique. This plan enables both Base Stations (BSs) and satellites to select files from a file library and subsequently store them in their respective local cache pools. The primary benefit of implementing this approach is that it significantly enhances the system's energy efficiency.

   \item Finally, we conclude by comparing our suggested optimization framework to benchmark algorithms in order to assess its effectiveness. Regarding energy efficiency, user satisfaction, and throughput, the experimental results show that our approach performs better than the other algorithms.
\end{itemize}

\begin{figure*}[!t]
	\centering
      \includegraphics[width=0.70\linewidth]{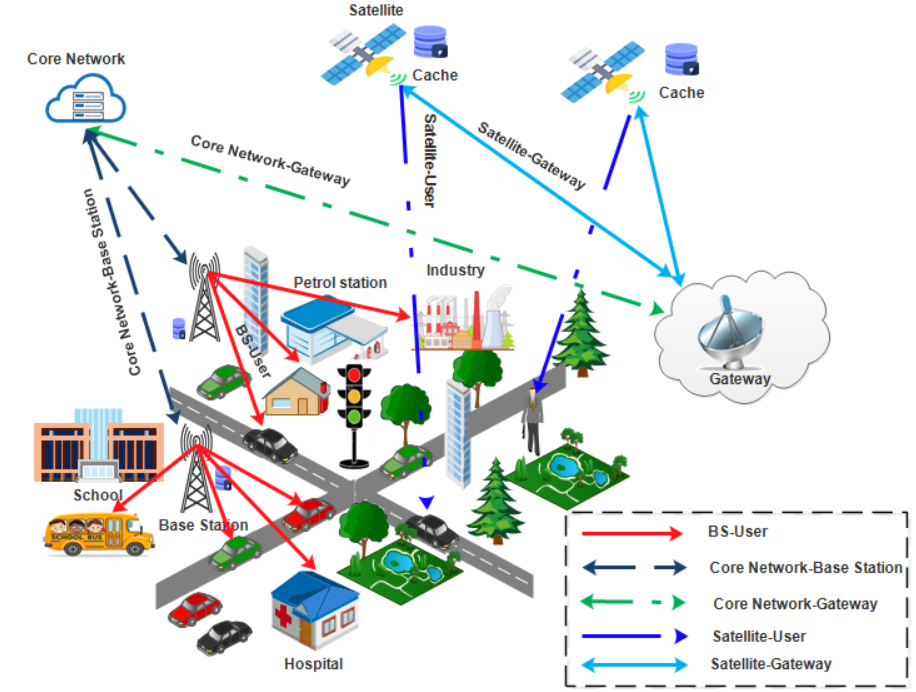}
	\caption{System Model}
	\label{fig:SM}
\end{figure*}
\section{System Model}
Motivated by the concept of integrated terrestrial non-terrestrial networks in \cite{liao2020distributed,gao2021sum,shaat2017joint,lagunas2019power,ji2020popularity,9992172}\footnote{Please note that integrated terrestrial and non-terrestrial networks are an emerging research area in academia and industry. The terrestrial networks are well-established and deployed. The non-terrestrial networks are expected to be deployed by the end of 2030. However, integrating both networks involves several challenges that need to be tackled.}, our study considers an integrated terrestrial-satellite network providing joint services to ground users, as shown in Figure 1. The network comprises a set of $M$ base stations (BSs) on the ground, denoted by $\mathbb{B}$, and $K$ low-orbit satellites, represented by $\mathbb{K}$. The users are divided into two groups: $N_b$ users are served by ground BSs, while the remaining $N_s$ users are served by satellites.

To mitigate user interference, we utilize a Non-Orthogonal Multiple Access (NOMA) scheme for BS-connected users \cite{khan2020secureMC}. This approach employs successive interference cancellation (SIC) at the receiver and superposition coding at the transmitter, allowing for sequential detection, demodulation, and interference cancellation. Users associated with a single BS are clustered into NOMA groups, where decoding priority is given to those with superior channel information, reducing interference from the users with higher path loss
\par
Moreover, the user association is a critical element of the proposed integrated terrestrial and satellite communication network, which enables users to connect to either a base station or a satellite during each time slot $t$. To represent these connections, binary variables $\alpha_n^m(t)$ and $\alpha_n^k(t)$ are used to indicate whether the $n^{th}$ user is connected to the $m^{th}$ base station or the $k^{th}$ satellite at time $t$. These binary variables play a crucial role in optimizing network performance, as they directly impact the quality of service for users and the overall network capacity. Optimization objectives such as minimizing interference, maximizing throughput, or balancing network load can be achieved by adjusting these binary variables.
\par
{Furthermore, the SINR of the $n^{th}$ user connected to the $m^{th}$ BS in a given time $t$ is calculated using the implemented NOMA scheme \cite{9712216}:
\begin{equation}
\small
\gamma_n^m(t)=\frac{\alpha_n^m(t)\left|g_n^m(t)\right|^2 p_n(t)}{I_{n'}^{m}(t)+I_{n'}^{m'}(t)+I_k(t)+N_o},
\end{equation}
Where in the above equation, each user has a transmit power $p_n(t)=\beta_n( p_b^{max}/N_b^1)$, which depends on a power control factor $\beta_n(t)$, the maximum power available at the base station $p_{b}^{max}$, and the number of users a single base station can serve $N_b^1$. Following that, the channel between the $n^{th}$ user  to $m^{th}$ BS is represented by $g_n^m(t)=\sqrt{\hat{g}_{n,m}d_{n,m}^{-\xi}}$, where $\hat{g}_{n,m}$ denotes Rayleigh fading coefficient, $d_{n,m}$ is the distance and $\xi$ represents the pathloss exponent \cite{khan2021noma}. {Practically, the distance $d_{n,m}$ can be calculated using the associated localization method, which involves cooperative positioning among multiple aircraft in cellular networks, as discussed by the authors in \cite{liu2021uav, liu2021v2x, liu2023management}}. Similarly, interference in the network is caused by users at the same base station and users in other base stations. From the satellite, users are represented by $I_{n'}^m$, $I_{n'}^{m'}$ and $I_k$ respectively. Whereas $N_o$ represents the noise spectral density. Moreover, the interference caused by users in the same base station is calculated based on the channel gains between users and is represented as $I_{n'}^m(t)=\sum_{{n'}\ne n} \alpha_{n'}^m(t)\left|g_{n'}^m(t)\right|^2 p_{n'}(t)$. Hence, the interference caused by the ${n'}^{th}$ users forms ${m'}^{th}$ base stations are determined by summing over all active users in other base stations, represented as $I_{n'}^{m'}(t)=$ $\sum_{{m'} \neq m} \sum_{{n'}=n}^{N_b^{m'}} \alpha_{n'}^{m'}(t)\left|g_{n'}^{m'}(t)\right|^2 p_{n'}(t)$. Moreover, the interference caused by the user connected to the satellite is determined by the channel gain between the user and the base station, represented as $I_k(t)=\sum_{k=1}^K \sum_{n=1}^{N_s} a_n^k(t)\left|h_n^m(t)\right|^2 p_n^k(t)$. The transmission power of the $n^{th}$ satellite user is denoted as $p_n^k(t)$.}
\par
Hence, the SINR of $n^{th}$ satellite user can be expressed as \cite{9375493}: 
\begin{equation}
\gamma_n^k(t)=\frac{\alpha_n^k(t)\left|h_n^k(t)\right|^2 p_n^k(t)}{I_n^m(t)+I_{n'}^{k}(t)+I_{n'}^{k'}(t)+N_o},
\end{equation}
{where in the above expression, $h_n^k(t)$ represents the block faded channel between the user $n$ and their associated $k^{th}$ satellite such that $h_n^k(t)=\hat{h}_{n,k}^{j\pi\vartheta}$,, where $\hat{h}_{n,k}$ denotes the complex-valued channel coefficient, $\vartheta$ is the Doppler shift while $j=\sqrt{-1}$. Moreover, the transmission power control  of the $n^{th}$ satellite user, $p_n^k(t=\beta_n( p_k^{max}/N_s^2))$, is determined by the power control factor $\beta_n(t)$ and the maximum available power $p_k^{max}$. Each satellite can serve up to $N_s^2$ users. Interference in the network is caused by users in base stations and other satellite users, which are denoted by $I_n^m(t)$, $I_{n'}^{k}(t)$ and $I_{n'}^{k'}(t)$, respectively. Moreover, the interference from users associated with the BSs can be calculated as $I_n^m(t)(t)=\sum_{m=1}^M \sum_{n=1}^{N_b^1} \alpha_n^m(t)\left|g_n^k(t)\right|^2 p_n^m(t)$. Similarly, the interference from other users from the same satellite as well as from other satellite users can be calculated as $I_{n'}^{k}(t)=\sum_{n' \ne n}^{N_s^2} \alpha_{n'}^k(t)\left|h_{n'}^k(t)\right|^2 p_{n'}^k(t)$ and $I_{n'}^{k'}=\sum_{{k'}\ne k} \sum_{n'\ne n} \alpha_{n'}^{k'}(t)\left|h_{n'}^k(t)\right|^2 p_{n'}^k(t)$. Here, $n$ is the index of the current satellite user, $M$ is the total number of base stations in the network, and $K$ is the total number of satellites in the network. Let us assume large and small scale fading \cite{khan2022rateconf}; then the complex-valued channel coefficient  can be defined as:
\begin{align}
\hat{h}_{n,k}= \sqrt{G_kG^k_{n} \Big(\frac{c}{4\pi f_c d^k_{n}}\Big)^2},
\end{align}
where $c$ defines the speed of light, $f_c$ states the carrier-frequency, $d^k_{n}$ represents the distance from the satellite, $G^k_{n}$ describes the antenna gain at receiver, and $G_{k}$ shows the antenna gain at the satellite. It is important to mention that the antenna gain of the satellite $G_{k}$ generally depends on the radiation pattern and the ground terminal location. It can be written as:
\begin{align}
 G_{k} =   G_{max}\left[\frac{J_1(\Lambda^k_{n})}{2\Lambda^k_{n}}+36\frac{J_3(\Lambda^k_{n})}{(\Lambda^k_{n})^3}\right]^2,
\end{align}
where $G_{max}$ is the maximum gain at the beam of satellite, $\Lambda^k_{n}=2.07123\sin{(\theta_{l,\iota})}/\sin(\theta_{3dB})$ such that $\theta^k_{n}$ shows the angle between the ground terminal and the satellite for any given location, where the 3 dB loss related to the satellite beam, which is given by $\theta_{3dB}$. Further, $J_1$ and $J_2$ represent the Bessel functions of the first and second orders, respectively.} 
\par
Following that, the $n^{th}$ users energy efficiency at time slot $t$ can be expressed as follows:
\begin{equation}
\label{eq3}
\small
\begin{aligned}
 \Psi_n(t)=\sum_{m=1}^m \alpha_n^m(t) \frac{R_n^m(t)}{p_n^m(t)}
 +\sum_{k=1}^K \alpha_n^k(t) \frac{R_n^k(t)}{p_{s, m}(t)}, \forall n, l.
\end{aligned}
\end{equation}
{Moreover in \eqref{eq3}, $R_n^m(t)=\log _2\left(1+\gamma_n^m(t)\right)$, $R_n^k(t)=\log _2\left(1+\gamma_n^k(t)\right)$ reprsent the achievable rates. Similarly, the proposed system model facilitates the file retrieval process for network users by utilizing cache pools in base stations and satellites. To request a file, users can access a file library $U = \{1, \dots, U\}$, and the cache pool size is fixed and based on the number of files and their size. Specifically, each base station and satellite have a cache pool of size $M_u < U$ and $M_s < F$, respectively, where the base station can store $M_u \times s$ bits of files, and each satellite is capable of storing $M_s \times s$ bits of files.}
\par
Similarly, when a $n^{th}$ puts in a request for a file, to address this, the system first checks if the requested file is available in the pool of the local cache of the base station. Let $L_m$ denote the set of files cached at base station $m$. If the requested file is in $L_m$, the file can be transmitted to the user, and the power consumed during this process is represented by $p_{m,r}(t)$. If the file is not available in $C_m$, the user looks for the file in the core network, which incurs a power consumption of $p_{l,r}(t)$.
\par
{Similarly, if the requested file is available in the satellite cache pool, In that case,  the user retrieves the file directly, and the power consumed during this process is represented by $p_{k,r}(t)$, where $L_k$ denotes the set of files cached at satellite $lk$. If the requested file is unavailable in $L_k$, the user's request is forwarded toward the ground gateway, and a file is accessed from the core network, resulting in a power consumption of $p_{l,r}(t)$.}
\par
The caching system provides benefits in terms of reducing time delays and alleviating power consumption by prioritizing locally cached files over files that need to be retrieved from the core network. The caching gain is dependent on whether the requested file is found in the cache pool (local) or not.
\par
{Similarly, the $J_m(t)$ indicates whether the local cache device satisfied the $mth$ BS user's file request at time $t$.
\begin{equation}
J_n(t)= \begin{cases}1, & \text {Request Satisfied } \\ 0, & \text {Otherwise} .\end{cases}
\end{equation}
Files in the system are assumed to follow the Zipf distribution, and their popularity affects the caching effect. In this regard, a generalized Zipf distribution is used in the system to estimate values of $\varepsilon$ ranging from $0.56$ to $0.83$, respectively \cite{zhang2019double}.
\begin{equation}
y_m=\frac{1 / u^{\varepsilon}}{\sum_{u=1}^U (u^{\varepsilon})^-1}, \forall u .
\end{equation}
The reward for caching deployment is given by reducing the time delay can be expressed as follows: 
\begin{equation}
x_n(t)=J_n(t) {\text {C}_n^s}{T_n^{-1}},
\end{equation}   }
The system model represents the time delay $T_n$ associated with downloading requested content for the user $n$  via a back-haul link. Moreover, the size of the cache file is denoted as $s$, and $C_n$ represents the content requested by the $n^{th}$ user. When the requested content is available in the local cache, it can be directly obtained from there, reducing the time delay.
\par
Likewise, it's worth examining the advantages of using a satellite cache in terms of minimizing latency and transmitting data that is cached, thereby improving overall performance, which can be expressed as: 
\begin{equation}
x_n^k(t)=J_m(t) C_n^k (T_m^k)^{-1}
\end{equation}
The time required for the $n^{th}$ user to download the requested content through the back-haul link is $(T_mk)$, indicating a delay in the process.
\par
The cache policy's effectiveness is evaluated by examining the hit rate for the cache, which represents the proportion of requests from users that are successfully fulfilled. To calculate the cache hit rate for a given duration of time $t$, the following formula can be used:
\begin{equation}
\Omega(t)=\frac{\sum_{n=1}^N J_n(t)}{N} .
\end{equation}
We will use the notation $P(t)$ to represent the total amount of power used by all BS users at $t$ time.
\begin{equation}
P(t)=p_n(t)+J_n^{'}(t) p_{l, r}(t)+J_n(t) p_{n, r}(t),
\end{equation}
Where $J_n^{'}(t)=\left(1-J_n(t)\right)$. In our system model, the power consumption for both BS and satellite users can be expressed using the variables $P(t)$ and $P_k(t)$, respectively. Power consumption for BS users includes both the transmission power of user $n$, denoted as $p_n(t)$ and the power consumption for data retrieval. The latter is further subdivided into two parts: $p_{n,r}(t)$ for data retrieval from the  BS cache as well as $p_{l,r}(t)$  from the core network via the back-haul link, respectively. On the other hand, for users connected to the satellite, we utilize the variable $P_k(t)$ to represent the total power consumption, including the power consumption for transmission and data retrieval.
\begin{equation}
P_k(t)=p_{k, n}(t)+\left(1-J_n(t)\right) p_{k, l, r}(t)+J_n(t) p_{k, n, r}(t),
\end{equation}
{Various factors determine the power consumption of a user in a satellite network. The transmit power $p_k^n(t)$ required for the user $n$ to communicate with the satellite, as well as the power consumed during data retrieval from the satellite's cache $p_{k, n, r}(t)$ or the core network via the Gateway Station $p_{k, l, r}(t)$, are examples. We combine the BS cache with the satellite to determine how well the satellite network uses energy. This allows us to estimate $n^{th}$ user energy efficiency over $t$. Energy efficiency is a crucial metric for figuring out how well the satellite network works and how to improve its design to give users the most benefits while using the least power. It can be expressed as follows:
\begin{equation}
\label{e11}
\begin{aligned}
 \Psi_n(t)\!= \!\!&\sum_{m=1}^M \alpha_n^m(t) \frac{R_n^m}{p_n(t)\!+J_n^{'} p_{l, r}(t)+J_n(t) p_{n, r}(t)} \\
& +\sum_{k=1}^K \alpha_n^k(t) \frac{R_n^k)}{p_n^k(t)} .
\end{aligned}
\end{equation}
Where in \eqref{e11}, $J_n^{n'}1-\!J_m(t)$.  }
\section{Problem Formulation}
  This work aims to maximize the system's overall energy efficiency by optimal allocation of resources, e.g., transmission power, user association matrix, and cache layout. Furthermore, it seeks to minimize the system's overall energy consumption while maintaining high performance and quality of service. The ultimate goal is to achieve an optimal balance between energy efficiency and system performance by finding the most efficient way to allocate resources while meeting operational requirements. The associated constraint represents the restriction that each user is limited to a single BS or satellite during a specific period and can be expressed as follows: 
\begin{equation}
\small
\sum_{m=1}^M \alpha_n^m(t)+\sum_{k=1}^K \alpha_n^k(t) \leq 1, \quad \forall n.
\end{equation}

A maximum power constraint exists for each user associated with a BS or a satellite. The transmission power limit for users associated with a BS is given by:

\begin{equation}
\label{c4}
\small
p_n(t) \leq \frac{p_b^{\max }}{N_b^1}, \quad \forall n.
\end{equation}

The transmission power limit for users associated with a satellite is given by:

\begin{equation}
\label{c5}
\small
p_n^k(t) \leq \frac{p_k^{\max}}{N_s^2}, \quad \forall n.
\end{equation}

It is important to remember that the QoS limitations of each BS and satellite limit the maximum number of users they can accommodate. The maximum number of users for a specific BS is:
\begin{equation}
\small
\sum_{n=1}^{N_1^b} \alpha_n^m(t) \leq N_1^b, \quad \forall n.
\end{equation}

The quantity of service constraint for a satellite is:

\begin{equation}
\small
\sum_{n=1}^{N_s^2} \alpha_n^k(t) \leq N_s^2, \quad \forall n .
\end{equation}

The following constraint shows that each user's power control factor is restricted to fall within $0$ to $1$.

\begin{equation}
\small
\beta_n(t) \in[0,1], \quad \forall n.
\end{equation}

The caching strategy of base stations and satellites is constrained by the capacity of their respective local caches. In addition, the size of user content requests is smaller than the available local storage capacity, which is still insufficient to accommodate the total size of all file libraries. The constraint that describes the limitation on the local cache capacity for BS and satellite is as follows:

\begin{equation}
\begin{aligned}
C_n \leq M_f \leq U, \
C_n \leq M_s \leq U .
\end{aligned}
\end{equation}
Based on the aforementioned objective and constraints, the optimization problem can be expressed mathematically. One possible formulation is as follows:
\begin{subequations}
\small
\begin{align}
\max &\sum_{m=1}^M\sum_{m=1}^M \alpha_n^m(t) \eta_n^m
 +\sum_{k=1}^K \alpha_n^k(t) \psi_n^k. \nonumber\\
&\sum_{m=1}^M \alpha_n^m(t)+\sum_{k=1}^K \alpha_n^k(t) \leq 1, \quad \forall n. \\
& \sum_{n=1}^{N_1^b} \alpha_n^m(t) \leq N_1^b, \quad \forall n. \\
& \sum_{n=1}^{N_s^2} \alpha_n^k(t) \leq N_s^2, \quad \forall n . \\
& p_n(t) \leq \frac{p_b^{\max }}{N_b^1}, \quad \forall n. \\
&p_n^k(t) \leq \frac{p_k^{\max}}{N_s^2}, \quad \forall n. \\
& \beta_n(t) \in[0,1], \quad \forall n.\\
& C_n \leq M_f \leq U,\\
&C_n \leq M_s \leq U .
\end{align}
\end{subequations}

Where, $\eta_n^m=\frac{R_n^m}{p_n(t)\!+J_n^{'} p_{l, r}(t)+J_n(t) p_{n, r}(t)}$ and $\psi_n^k=\frac{R_n^k)}{p_n^k(t)} $.
The optimization problem mentioned above appears to be a mixed-integer nonlinear optimization problem (MINLP). This is because it involves both integer and continuous variables and nonlinear constraints such as power constraints in constraints \eqref{c4} and \eqref{c5}. Additionally, the objective function is also nonlinear. MINLPs are known to be challenging to solve as they combine the computational difficulties of both nonlinear and integer optimization problems.
\section{Optimizing Terrestrial-Satellite Network Efficiency with Multi-Agent DRL (MADDPG)}
{This section outlines a MADDPG approach for enhancing the integrated terrestrial-satellite NOMA communication network. The main aim of this work is to maximize the objective function value by determining the optimal allocation of resources, e.g., transmission power control, the design of the cache, and the allocation of users. Therefore, to achieve this, we suggest two MADDPG algorithms that concentrate on different subproblems. To ensure peak performance, both algorithms carefully choose the agents.}
\subsection{Reinforcement Learning}
The objective of reinforcement learning (RL), a type of machine learning, is to teach an agent how to interact with the environment to maximize a cumulative reward signal. RL does not always need a dataset to learn from, unlike supervised learning, which necessitates a labeled dataset. Instead, an RL agent can learn by interacting with its surroundings, getting feedback through rewards or penalties, and then changing its behavior.
\par
The agent participates in the RL process by acting in the world and receiving feedback as a reward signal. In order to raise its expected cumulative reward, the agent modifies its policy, a function that links states to actions. Modifying the parameters of the decision-making model is part of the learning process, which involves updating the policy.
\par
The agent continuously refines its behavior through trial and error, one of RL's advantages. The agent experiments with various actions in the environment, evaluates the rewards that result, and then modifies its strategy to carry out more actions that produce greater rewards. This cycle repeats until the agent discovers a course of action that maximizes its cumulative reward.
\subsection{Enhancing NOMA Networks with MADDPG}
The integrated terrestrial-satellite NOMA communication network comprises multiple agents, which makes it a complex multi-agent scenario. In such a scenario, the suggested MADDPG algorithm is the most suitable approach due to its flexibility in handling many agents. On the other hand, traditional single-agent reinforcement learning methods may encounter overfitting issues against competitors in dynamic and unstable environments. The objective function value in an integrated terrestrial and satellite NOMA communication network is maximized by using a Markov decision process (MDP), State space (S), action space (A), reward space, and transition probability space must all be defined. We can only find an effective solution if we accurately model the problem. In this configuration, each user acts as an agent by monitoring their immediate environment, selecting appropriate actions from the available action space, and carrying those actions out to satisfy the configuration's prerequisites. When a user completes all of their tasks, they are rewarded. Regardless of the algorithm used, in both analyzed approaches, agents, actions, states, and rewards had distinct definitions.
\subsubsection{Multi-Agent Reinforcement Learning for User Association and Power Control in NOMA Networks}
An integrated terrestrial-satellite network's energy efficiency optimization problem can be resolved using the MADDPG algorithm, detailed in Algorithm 1. These key concepts, which include agents, actions, states, and rewards in this algorithm, are defined below:
\par
\textbf{agent}: Every participant in an integrated terrestrial and satellite-enabled NOMA communication network is treated as a potential agent.
\par
\textbf{Action}: In the above-mentioned system design, every agent is assigned $2$ tasks specified by the action space $A_1={A_{11}, A_{12}}$. The first task, denoted as $A_{11}$, involves the user association process, establishing the relationship between agents and the base stations (BSs) or satellites. The $A_{11}$ action is represented by a vector $A_{11}={a_1^n(t), \ldots, a_M^n(t)}$, where each entry corresponds to the association decision of a particular agent. To represent the $A_{11}$ action discretely, the action space $A_{11}$ must be discretized. The second task, represented by a vector $A_{12}={\alpha_1(t), \ldots, \alpha_M(t)}$, involves the power control factor that each agent uses to determine its transmission power. In summary, each agent first selects the appropriate BSs or the satellites for the association using discretized user association action $A_{11}$ and subsequently determines its transmission power using the transmission power control factor $A_{12}$.
\par
\textbf{Reward}: Similarly, each user aims to optimize energy efficiency by taking appropriate actions. To evaluate the effectiveness of these actions, we define the reward for the $n$-th user at the current time slot $t$ as follows:
\begin{equation}
\operatorname{R}_1(t)\Psi_n(t).
\end{equation}
\par
\textbf{State}: In this system, the state space for each agent is determined by its observation of energy efficiency. Specifically, for the user $n$  in time $t$, the state is determined by comparing its energy efficiency with the previous time slot. If there is an improvement, then $S_{1 m}^{\Psi_n}$ is $\approx$1. Moreover, for the entire system state space is defined as $S_1={S_{11}^{\Psi_n}(t), \ldots, S_{1 N}^{\Psi_n}}$, where $S_{1 m}^{\Psi_n}$ represents the state for the user $n$ in time slot $t$.
\begin{equation}
S_{1 i}^{\Psi_n}= \begin{cases}1, & \text { if } R_1(t) \geq R_1(t-1) \\ 0, & \text { else }\end{cases}
\end{equation}
where $i$ represents the index of $n$-th user.

 \subsubsection{MADDPG-based Cache Optimization} 
 In order to allocate the resource optimally in the integrated terrestrial and satellite NOMA communication network, we employ Algorithm 1. Once this optimization is complete, we proceed with the optimization for the design of the cache of the BS as well as for the satellites using Algorithm 2. Due to the differences in the optimization goals, the actions, agents, rewards, and states in Algorithm 2 may differ slightly from those used in Algorithm 1 and can be stated as follows:

\textbf{Agent}: Algorithm 2 treats BSs or satellites as agents deciding which files to retrieve from the library.

\textbf{Action}: Every satellite or BS determines files to utilize from the files library for each time slot. This set of files, denoted as $A2=\{A {21}\}$, constitutes the local cache pool by combining the file libraries.
\textbf{Reward}: The system's goal is to maximize its energy efficiency, which is achieved by having each base station (BS) or satellite execute a set of operations. The number of agents in the system totals $M+K$. The overall energy efficiency of the users served by the $n$-th base station or satellite is the reward for the operations of the facilities that operate those base stations and satellites. To be more specific, the sum of the reward for the $n$-th piece of bad behavior that occurs during the time slot $t$ depends on the quantity that is denoted as:
\begin{equation}
\small
\begin{aligned}
&\operatorname{R}_2(t) {\Psi_n}\\= &\begin{cases}\sum_{n=1}^{N_b} {\Psi_n}, & n \in[1, N_b], \\ \sum_{n=1}^{N_s} {\Psi_n}, & n \in[M+1, M+K]\end{cases}
\end{aligned}
\end{equation}
\textbf{State}: In Algorithm 2, the satellite or base station (BS) is the agent responsible for optimizing the system's energy efficiency. Thus, $S2_{pi{\Psi_n}}$ is set to 1 if the reward for the $n$-th BS or the satellite time $t$ slot is greater than that in the $t-1$. The system's state space is denoted by $S_2=\{S_{21}^{\Psi_n}(t), \ldots, S_{2N}^{\Psi_n}(t)\}$. The quantity AAA is assigned to $S2_{iE E(t)}$.
\begin{equation}
S_{2 i}^{\Psi_n}(t)= \begin{cases}1, & \text { if reward } \\ 0, & \text { otherwise. }\end{cases}
\end{equation}
In order to achieve a greater degree of stability, the MADDPG algorithm can gather data concerning the activities of various other agents. The probability of the change will be discussed in more detail in the upcoming paragraphs.
\begin{equation}
\label{25} 
\small
\begin{aligned}
& P\left(s^{\prime} \mid s, a_1, \ldots a_N, X_1, \ldots, X_M\right)=P\left(s^{\prime} \mid s, a_1, \ldots a_N\right) \\
& =P\left(s^{\prime} \mid s, a_1, \ldots a_N, X_1^{\prime}, \ldots, X_M^{\prime}\right) .
\end{aligned}
\end{equation}
A MADDPG algorithm for an integrated terrestrial-satellite NOMA communication network maintains stability even when agents' policies are dynamically updated. Equation \eqref{25} shows the state transition probability, with $a_i$ representing the action taken by the agent and $s$ representing the current state. The network includes $M$ agents, each with a set of corresponding parameter values $w={\varpi_1, \ldots, \varpi_M}$ optimized to maximize returns via the MADDPG algorithm.

The MADDPG algorithm uses reinforcement learning to enable multiple agents to learn from their experiences and improve policies cooperatively. Each agent updates its policy based on observations and a centralized critic estimating the expected return. The critic considers all agents' experiences and policies, facilitating collaboration and performance improvement.

All $M$ agents' policy values are represented by $x={x_1, \ldots, \chi M}$, and each agent optimizes its policy using its unique set of parameter values $w$. The following equation determines the gradient of the objective function, which measures how the policy should be updated for each agent. ByAgentsan achieves better network performance and increases overall system capacity.
 by optimising individual policies\begin{equation}
\label{26}
\small
\nabla_{x_i} I\left(\chi_i\right)=E_{x, a \sim D}\left[\nabla_{x_i} \chi_i\left(\alpha_i \mid \omega_i\right) \nabla_{a_i} Q_i^X\left(x, a_1, \ldots, a_N\right)\right],
\end{equation}
The MADDPG algorithm utilizes two distinct neural networks, namely the actor network and the critic network, in order to optimize the performance of all the M agents in the integrated terrestrial and satellite NOMA  networks. At the same time, the observation and action spaces of the agents are denoted by $x$ and $a$, respectively, while the replay memory is represented by $D$. The actor-network in the algorithm selects actions by taking into account the policy, using a continuous action space to choose between $A_1$ and $A_2$. On the other hand, the critic network in the algorithm is used to evaluate the actions that need to be executed by updating the $Q$ function, which is represented by $Q_1^\gamma\left(x, a_1, \ldots, a_N\right)$ as shown in equation \eqref{26}.
\par
To update the networks, the policy network of the actor-network is updated using gradient descent based on equation \eqref{26} \cite{arulku7deep}. Meanwhile, the critic network updates the $Q$ function by minimizing the loss function $L\left(\omega_i\right)$, as illustrated in the following equation.
\begin{equation}
\small
y=r_i+\left.r Q_i^{u^{\prime}}\left(x^{\prime}, a_1^{\prime}, \ldots, a_N^{\prime}\right)\right|_{a_j^{\prime}=\mathrm{u}_j^{\prime}\left(o_j\right)} .
\end{equation}

\subsection{Algorithm Description}
This section presents two algorithms, Algorithm 1 for system resource allocation and Algorithm 2 for system cache design, based on the MADDPG algorithm \cite{li2021multi}. Before starting the algorithms, the neural network parameters and replay memory are initialized. The actor-network probabilistic-ally selects behaviours, while the other network, also known as the critic network used to evaluate the chosen behaviours. In contrast, the actor adjusts the probability of the selected behaviour upon the evaluation from the critic network. In iterative MADDPG  for terrestrial-satellite network, each agent is assigned an initial state and then observes its new state at each step of an episode, with energy efficiency improved at each instance. After successfully executing an action, each agent is rewarded and transitions to a new state. Both policy and exploration inform the agent's decision on what to do. Finally, these values are stored in memory for potential future replay.
 \begin{figure}[h]
	\centering
	\includegraphics[width=0.95\linewidth]{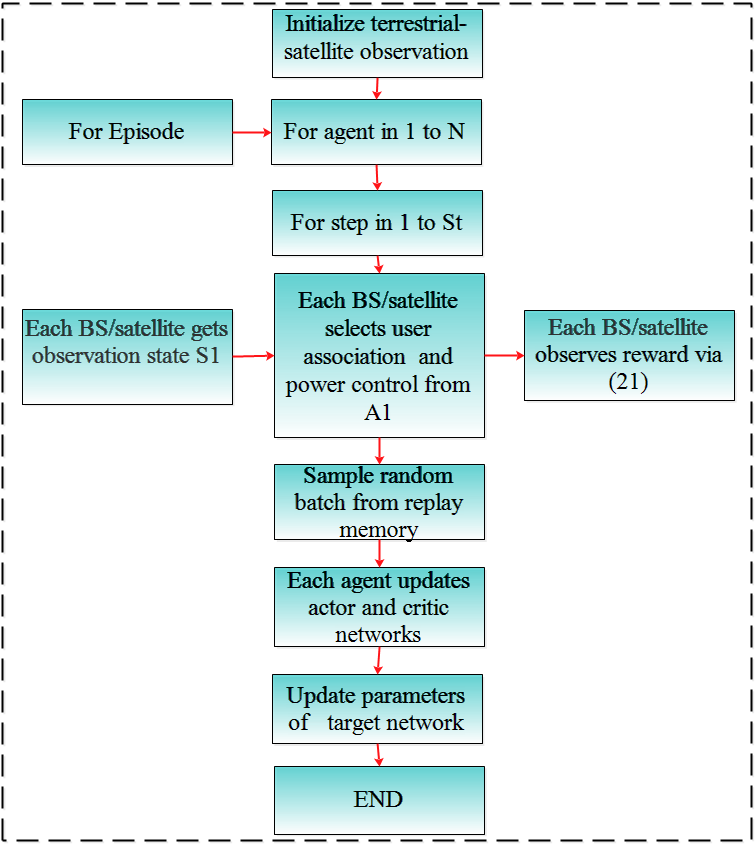}
	\caption{Algorithm 1 Flow Chart}
	\label{fig:2a}
\end{figure}

\begin{figure}[h]
	\centering
	\includegraphics[width=0.95\linewidth]{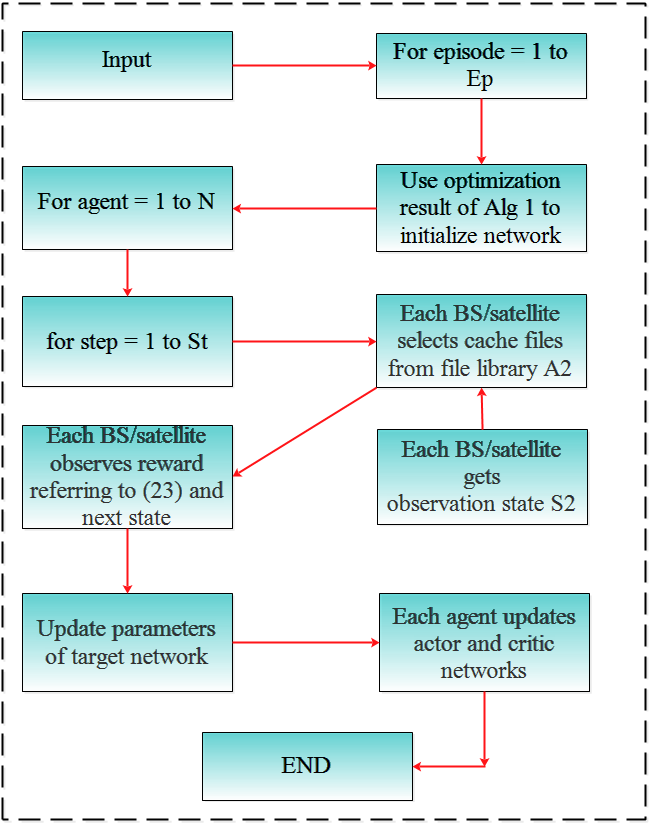}
	\caption{Algorithm 2 Flow Chart}
	\label{fig:2b}
\end{figure}

\begin{figure}[h]
	\centering
	\includegraphics[width=0.95\linewidth]{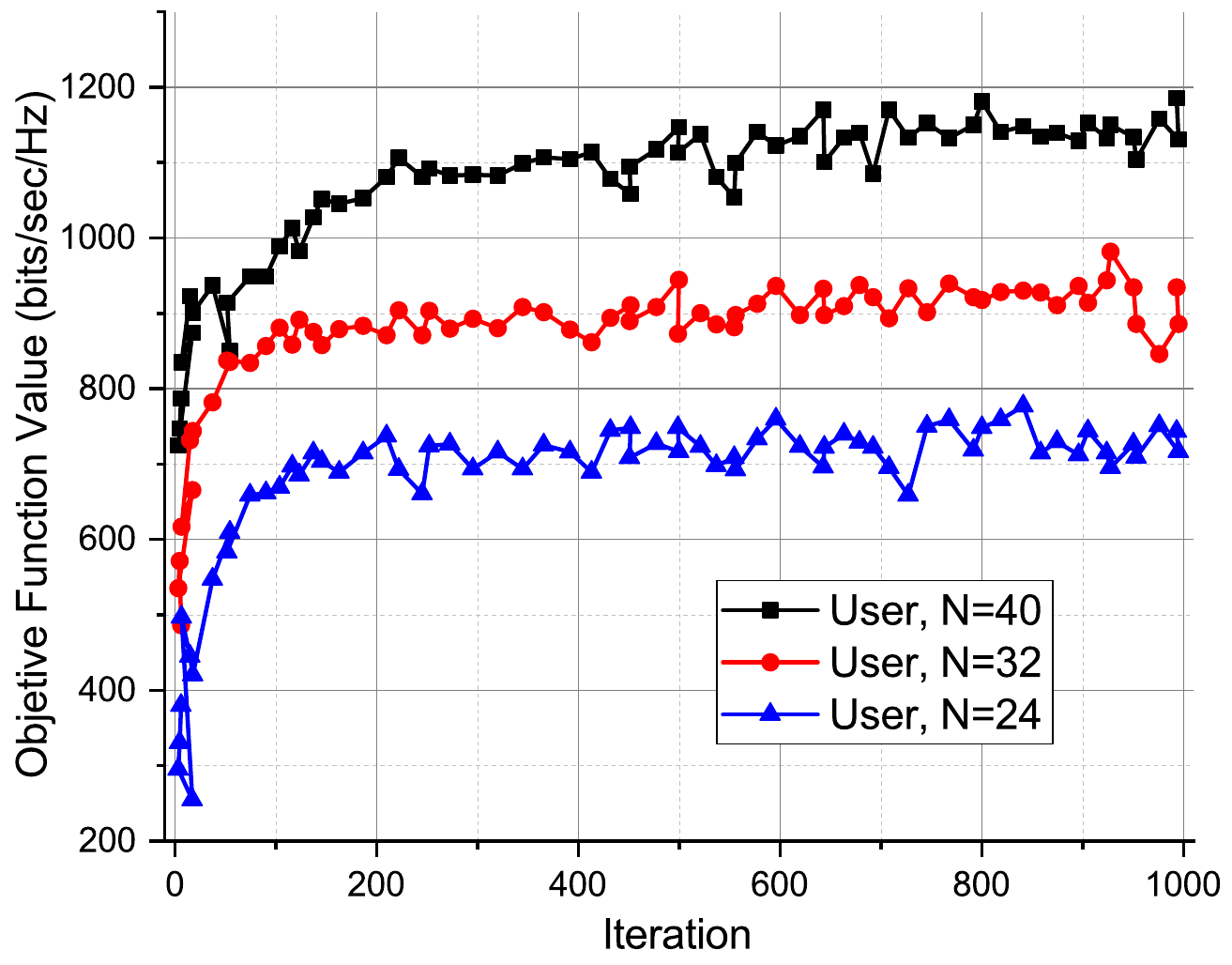}
	\caption{Algorithm convergence with varied numbers of Users}
	\label{fig:2}
\end{figure}

\begin{figure}[h]
	\centering
	\includegraphics[width=0.95\linewidth]{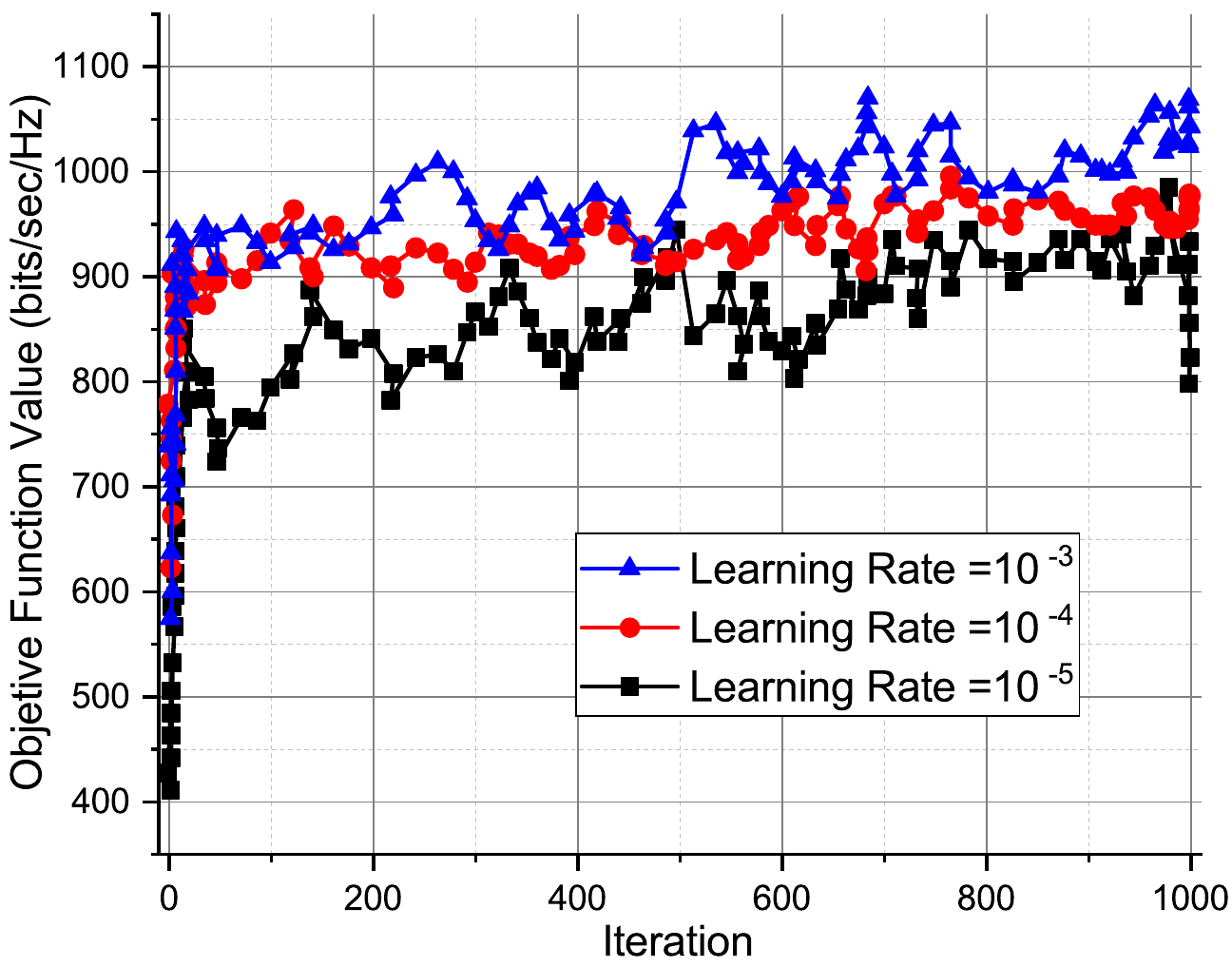}
	\caption{Effects of learning rate on Algorithm convergence}
	\label{fig:3}
\end{figure}

\begin{figure}[h]
	\centering
	\includegraphics[width=0.95\linewidth]{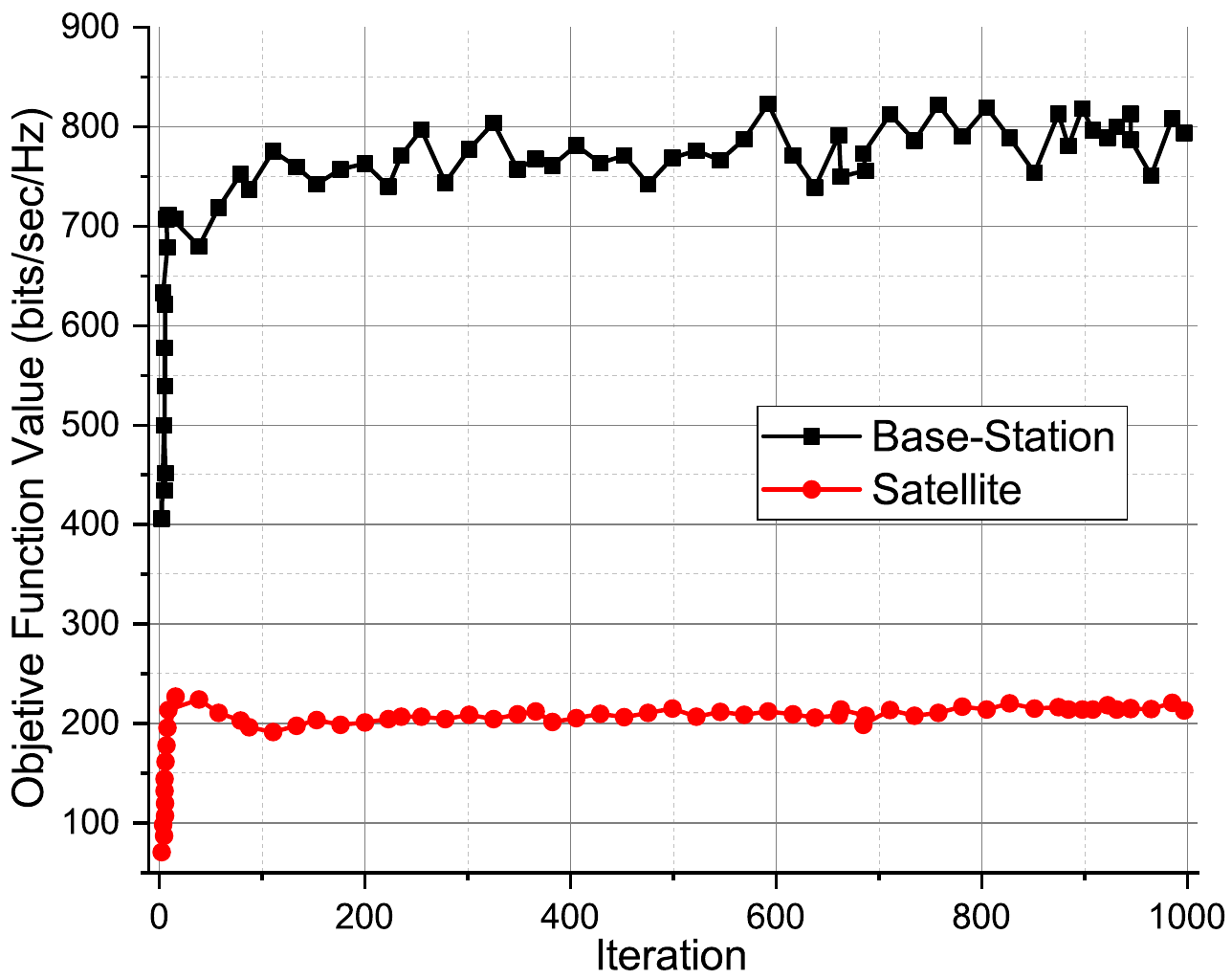}
	\caption{BS vs. Satellite users: a comparison of objective fuction value convergence}
	\label{fig:4}
\end{figure}

\begin{figure}[h]
	\centering
	\includegraphics[width=0.95\linewidth]{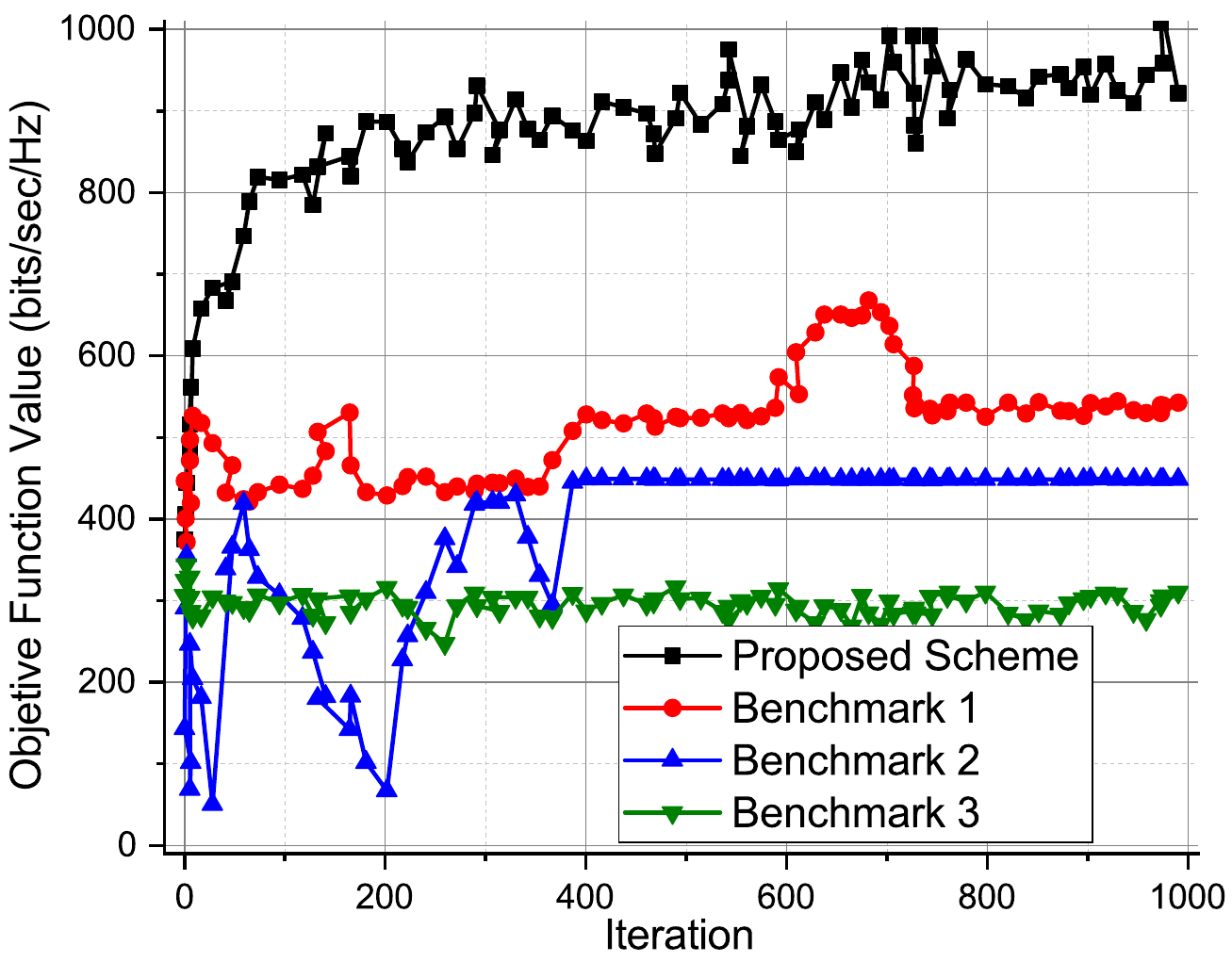}
	\caption{Proposed scheme versus benchmark algorithms: a comparison}
	\label{fig:5}
\end{figure}

\begin{figure}[h]
	\centering
	\includegraphics[width=0.95\linewidth]{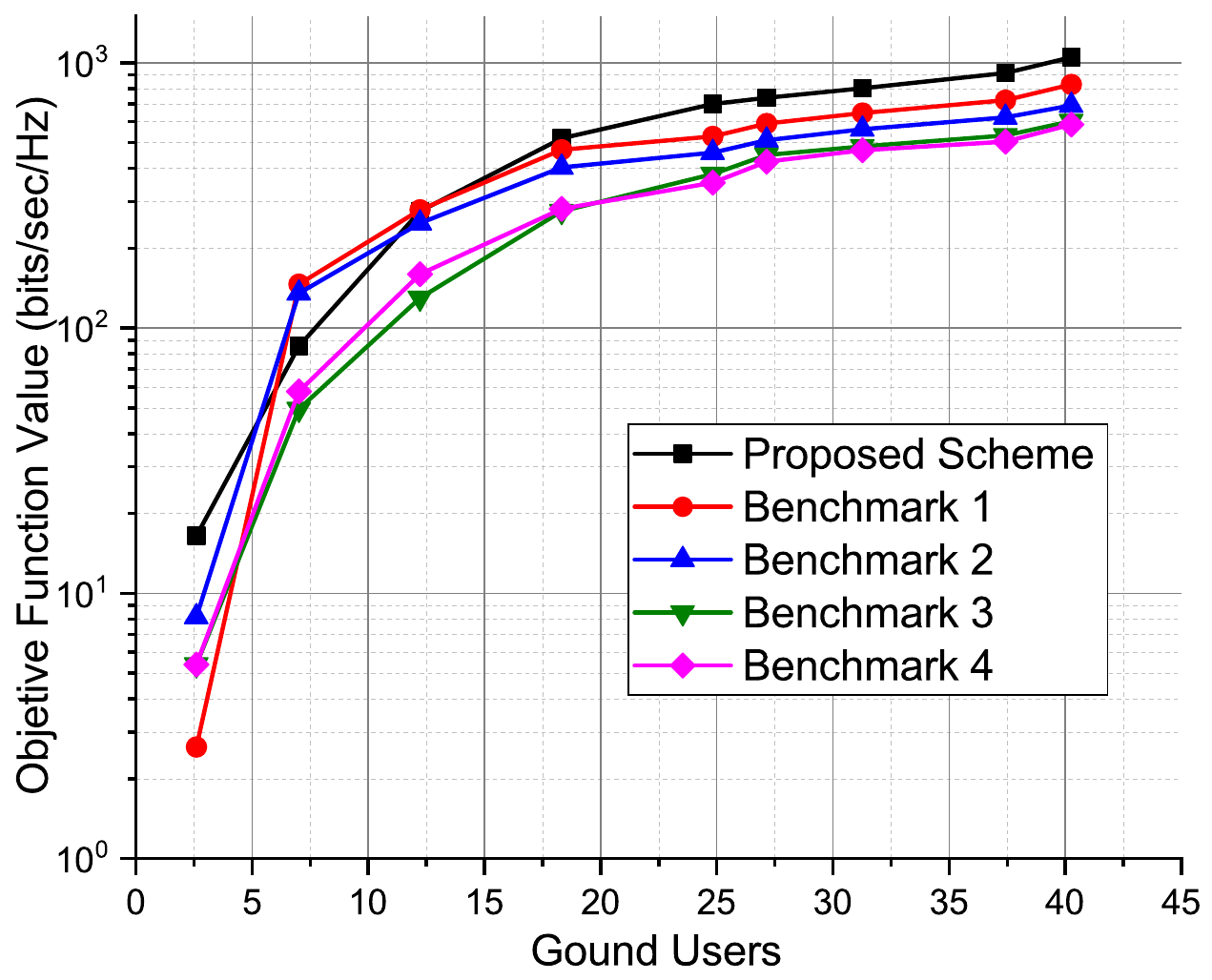}
	\caption{The impact of user count on objective function value of different algorithms}
	\label{fig:6}
\end{figure}

\begin{figure}[h]
	\centering
	\includegraphics[width=0.95\linewidth]{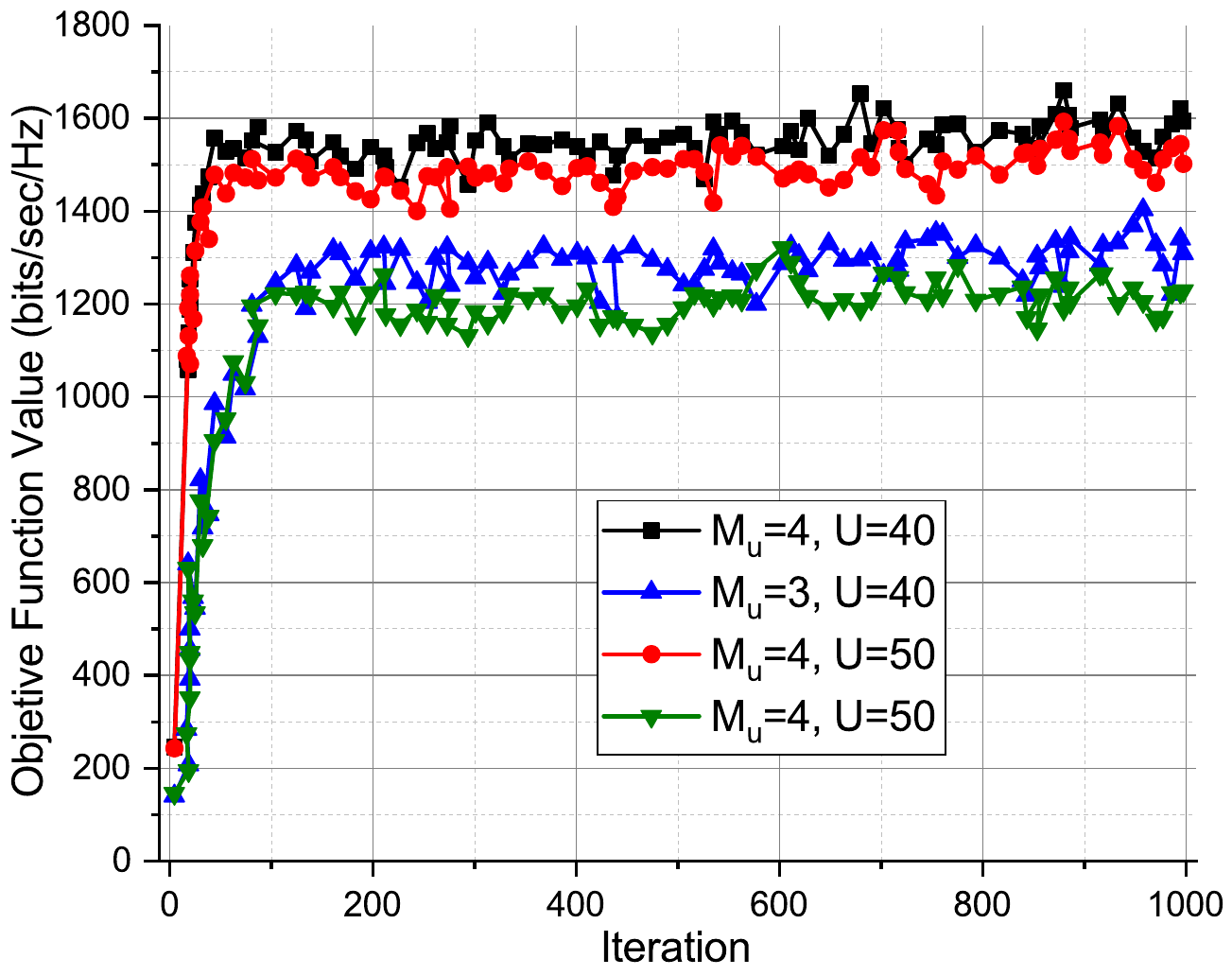}
	\caption{Cache Reward Proposed Scheme: convergence across capacity and library variations}
	\label{fig:7}
\end{figure}

\begin{figure}[h]
	\centering
	\includegraphics[width=0.95\linewidth]{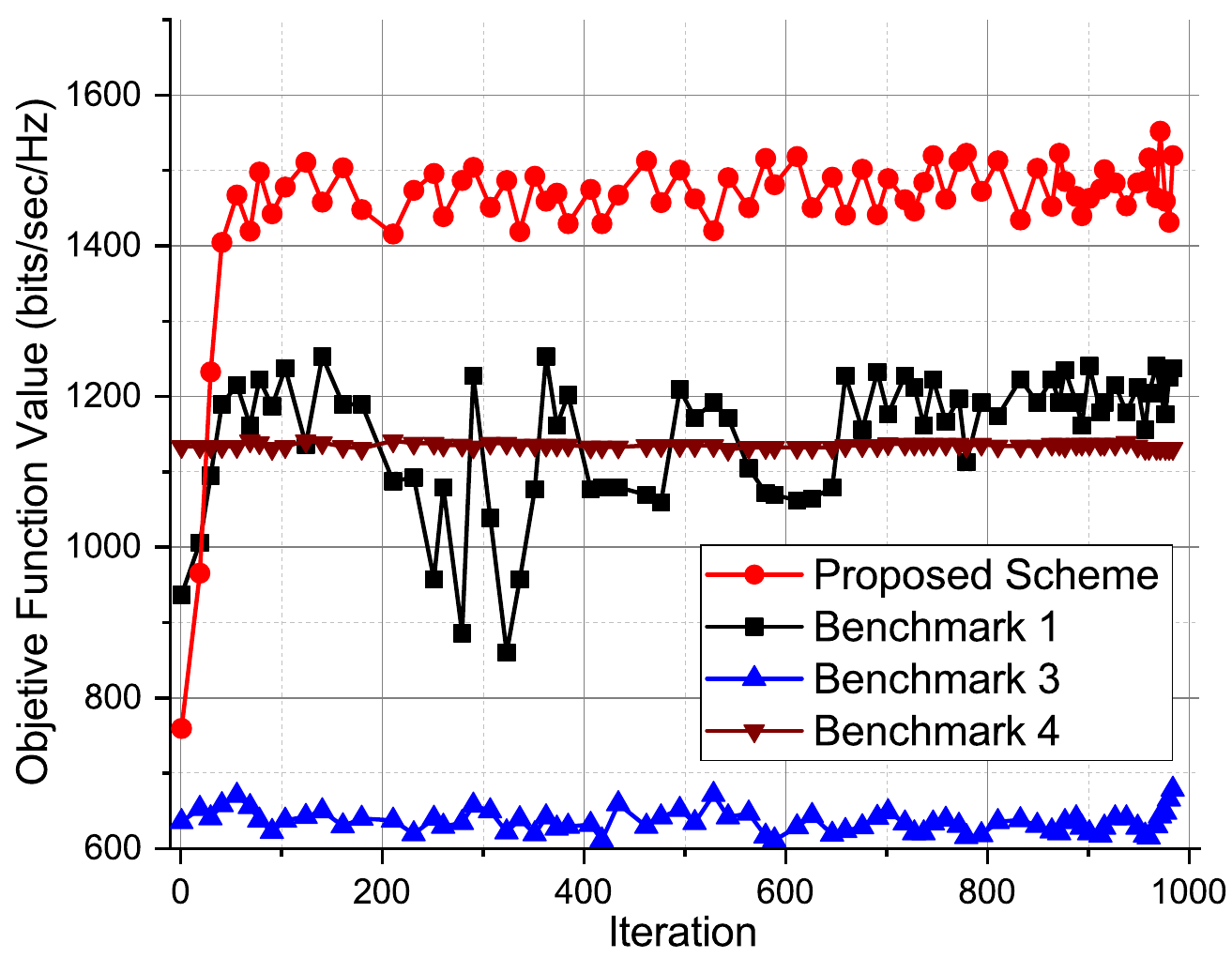}
	\caption{Proposed scheme versus benchmark algorithms: a comparison}
	\label{fig:8}
\end{figure}

\begin{figure}[h]
	\centering
	\includegraphics[width=0.95\linewidth]{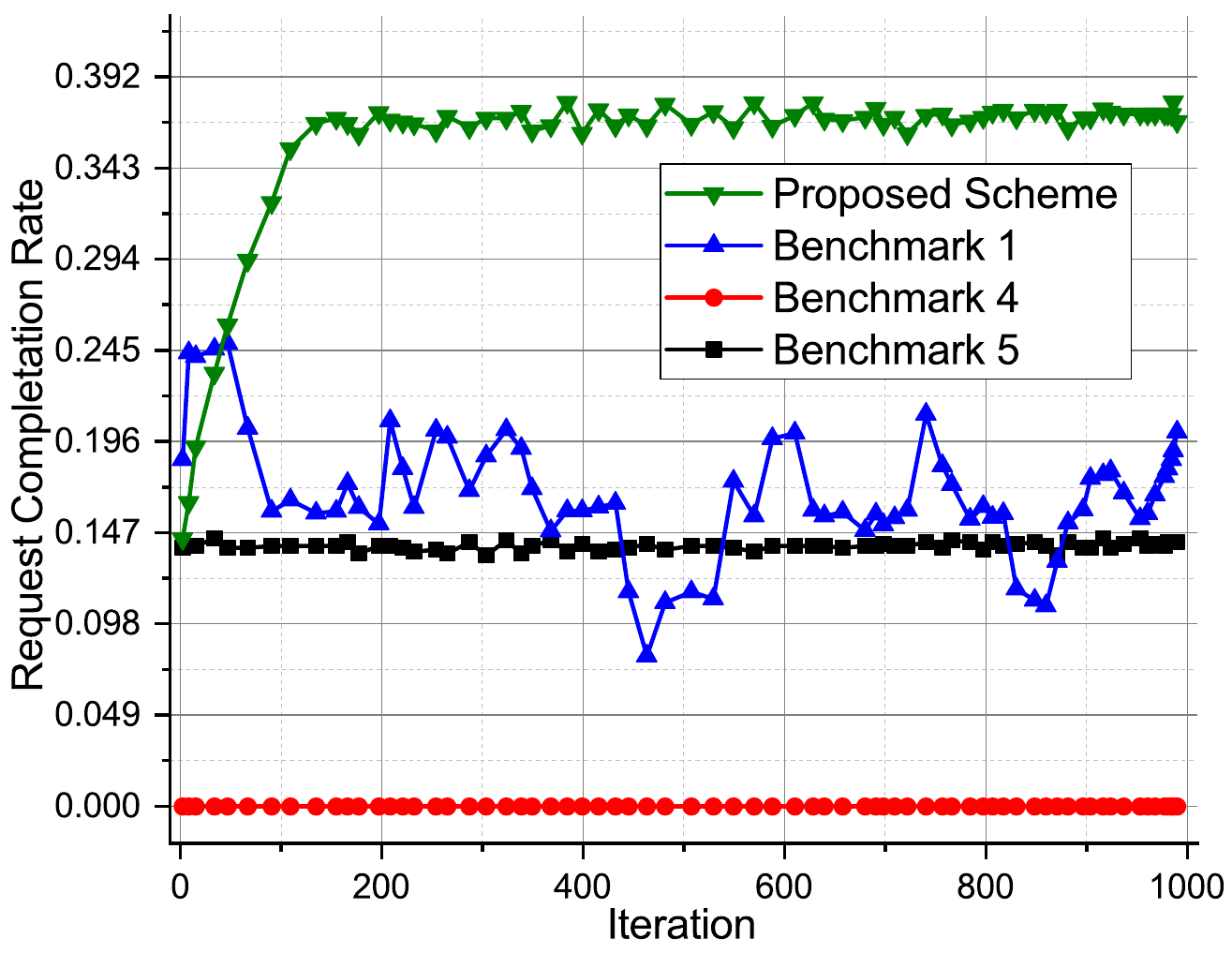}
	\caption{Proposed scheme compared to benchmark algorithms}
	\label{fig:9}
\end{figure}

\begin{figure}[h]
	\centering
	\includegraphics[width=0.95\linewidth]{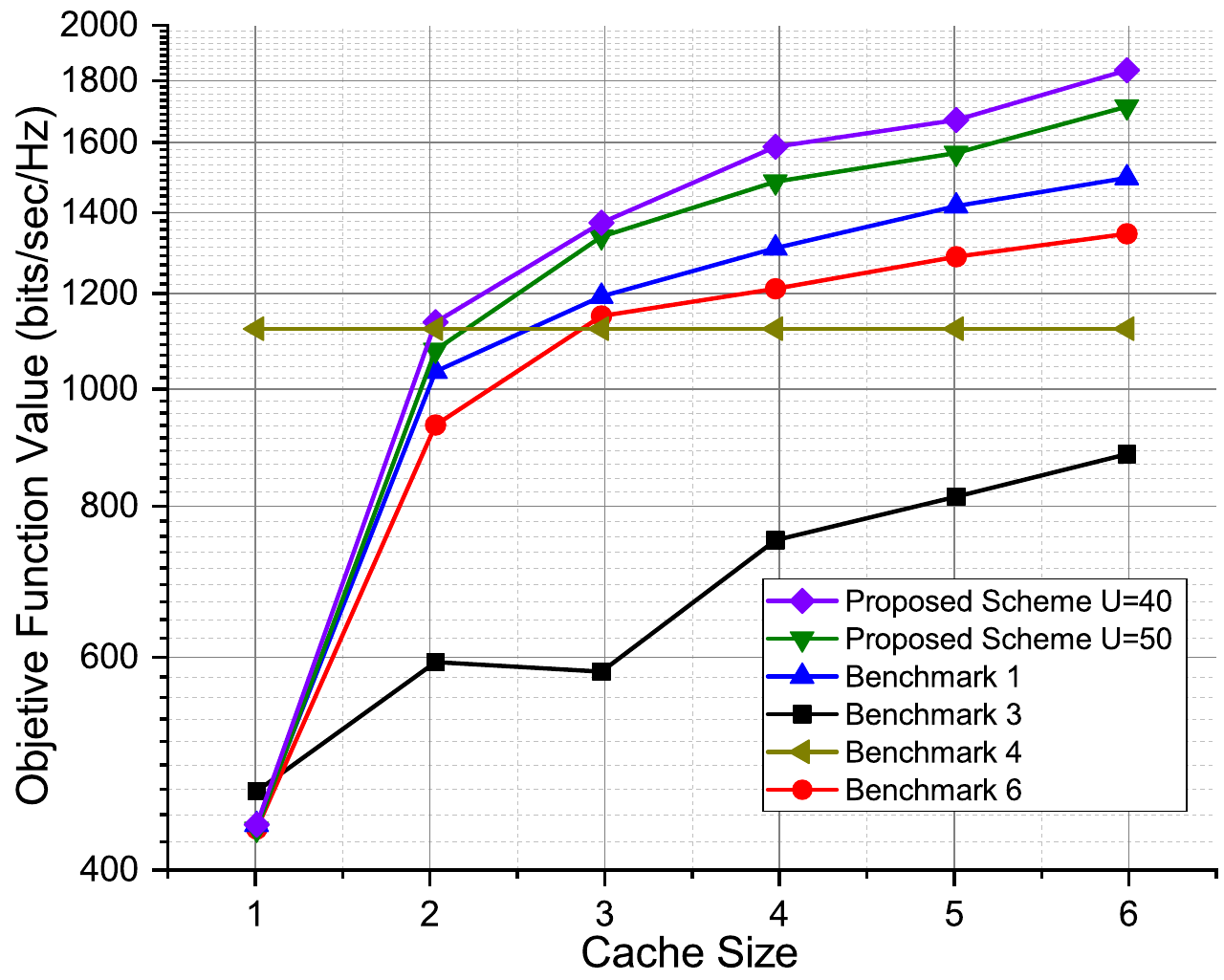}
	\caption{Proposed scheme and benchmark algorithms: a comparison across cache size variations}
	\label{fig:10}
\end{figure}

\begin{figure}[h]
	\centering
	\includegraphics[width=0.95\linewidth]{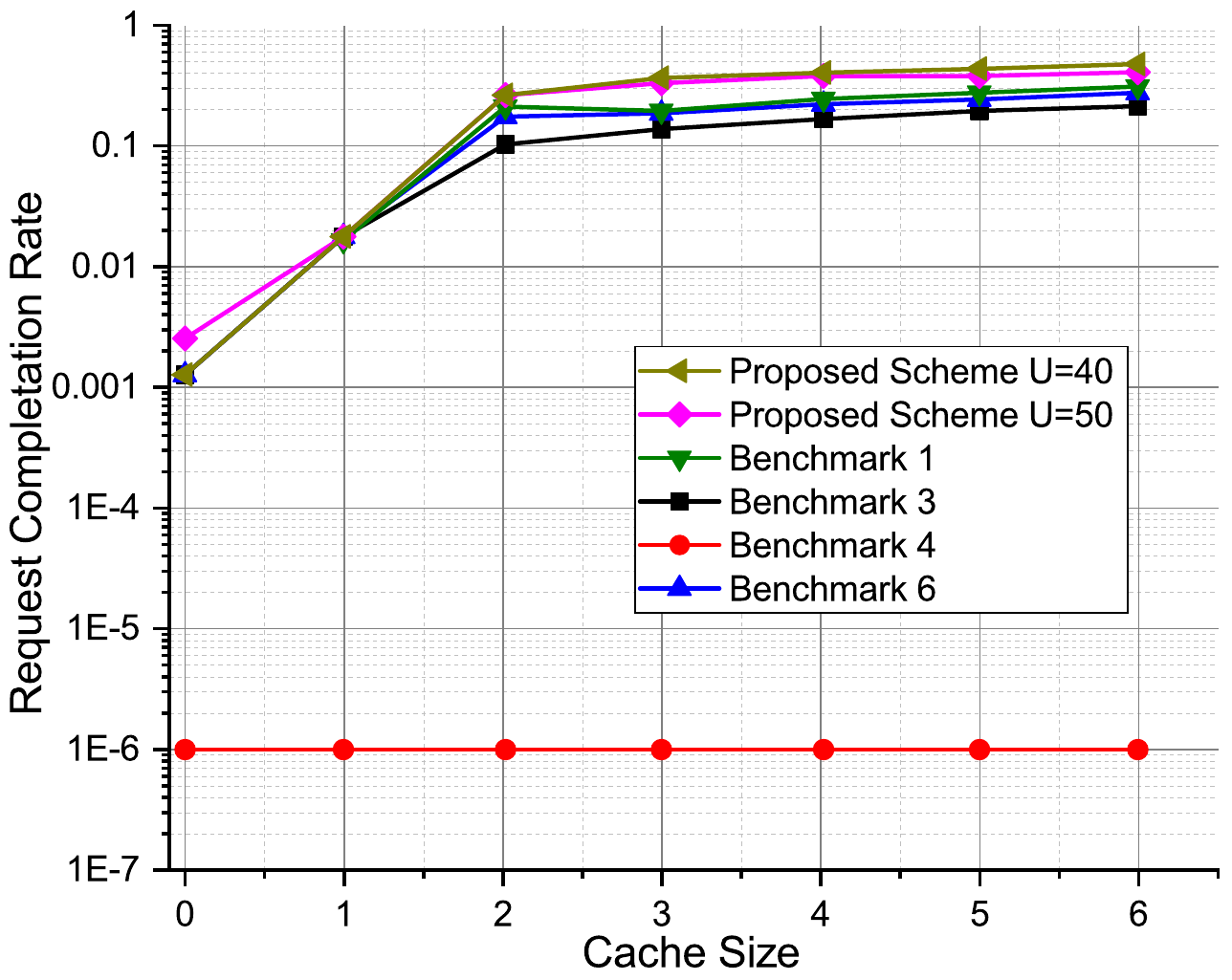}
	\caption{Evaluating proposed scheme and benchmark algorithms for Request Complementation rate under different cache sizes}
	\label{fig:11}
\end{figure}
\section{Results and Discussion}
In this study, we have employed a simulation-based approach to evaluate the performance of our proposed model\footnote{Given the unique and intricate nature of the network configuration under examination, it is not straightforward to directly compare it to previous studies on integrated terrestrial-satellite networks found in the existing literature. Therefore, we turn to the comparison of the proposed scheme with the different learning algorithms, including the deep reinforcement learning algorithm (DDPG), the Random Policy algorithm, the Genetic Algorithm (GA), and the Proximal Policy Optimization (PPO) algorithms.}. The experimental process involves the development of a synthetic/virtual simulation environment that closely represents the characteristics and behaviours of integrated terrestrial and satellite NOMA communication networks. The simulation environment is designed to mimic the network infrastructure, including base stations, satellites, users, and the communication channels between them. Various factors, such as user mobility, channel conditions, interference, and caching mechanisms, are considered to create a realistic simulation environment. We utilize MATLAB-based software tools and frameworks specifically designed for network simulations to implement the simulation environment. These tools provide capabilities for modelling and simulating the network components, implementing multi-agent deep reinforcement learning algorithms, and evaluating performance metrics. Furthermore, the experimental setup involves the deployment of our proposed multi-agent deep reinforcement learning model within the simulation environment. We carefully configure and tune the hyperparameters of the MADDPG algorithm, such as network topology, learning rates, exploration strategies, and reward functions, to ensure effective resource allocation and cache design. Our research primarily focuses on the algorithmic and methodological aspects of resource allocation and cache design in integrated terrestrial and satellite NOMA communication networks. Therefore, the experimental setup primarily involves the software-based simulation environment rather than physical or hybrid infrastructure. 
\par
The experimental environment encompasses specific configurations, including a network with 36 agents, consisting of 6 agents representing base stations and 2 agents representing satellites. The channel characteristics of the base stations follow a Rayleigh distribution, while the satellite parameters align with previous study findings. The file library denoted as U, has a capacity of 40, with 3 cache devices allocated for the base stations and a satellite cache with a capacity of 3. The user count and file content size is set at [1 2] bits. Power consumption parameters are also specified, covering data retrieval power consumption for different scenarios. The cache design optimization of the proposed scheme is conducted using Algorithm 2. Our simulation model utilises an Adam optimizer with the Rectified Linear Unit (ReLU) activation function. The learning rate is set to 0.001, the discount factor is 0.95, and the batch size is 10. The experiment is conducted over 1000 iterations, with each agent completing 100 steps per episode. While physical or hybrid infrastructure implementation considerations are significant in evaluating network performance, conducting large-scale physical experiments in integrated terrestrial and satellite networks can be complex, expensive, and challenging to scale. Hence, simulations provide a feasible and efficient approach to evaluate the performance of our proposed model and conduct extensive experiments under different scenarios.
\par
We assessed the performance of the proposed scheme in optimizing the system's objective function by evaluating its convergence with varying numbers of agents. Figure \ref{fig:2} showcases the convergence results of the proposed scheme with 24, 32, and 40 agents. To test the scheme's effectiveness in a more challenging environment, we increased the values of $N_b^1$ and $N_b^1$ to 5 while maintaining the network parameters at $N=24$, $M=6$, and $K=2$ for the 24-agent scenario. The curves in Figure \ref{fig:2} demonstrate that the proposed scheme consistently achieved a maximum reward of approximately 700 Iterations across all three agent numbers, demonstrating its strong convergence performance. The proposed scheme exhibited robust convergence behavior, even in complex terrestrial-satellite networks with multiple agents.

Furthermore, we conducted an analysis to assess the impact of different learning rates on the convergence performance of the proposed scheme. This experiment utilized network parameters with $M=32$, $N=6$, $S=2$, and $M_1=M_2=4$. As illustrated in Figure \ref{fig:3}, the curves exhibit varying convergence speeds and relative heights, depending on the learning rate. Notably, a higher learning rate resulted in a faster convergence point. This observation emphasizes the efficacy of the proposed scheme in optimizing the system's objective function across a diverse range of agent densities and learning rates.

The results depicted in Figure \ref{fig:4} demonstrate the convergence of both base stations (BSs) and satellites. Both graphs exhibit similar convergence rates, reaching their optimal states relatively quickly using the same experimental setup as in the previous figures. Figure \ref{fig:4} also highlights that users connected to BSs are more energy-efficient compared to satellite users. Satellite users converge to total energy efficiency of approximately 175 bits per joule per hertz, whereas BS users achieve a higher total energy efficiency of about 750 bits per joule per hertz. This efficiency disparity can be attributed to the improved channel conditions experienced by BS users.

To evaluate the optimization performance of the proposed scheme, we conducted experiments employing four different algorithms. Figure \ref{fig:5} illustrates the comparison between the proposed scheme and benchmark algorithms: the widely used deep reinforcement learning algorithm, DDPG, serving as a benchmark; the Random Policy algorithm, the Genetic Algorithm (GA), and the Proximal Policy Optimization (PPO) algorithm. The GA algorithm simulates natural evolution to identify the optimal solution, while the PPO algorithm is a newly developed policy gradient algorithm. The Random Policy algorithm randomly selects user collaboration and power control actions in each episode.

The performance of these algorithms in optimizing the system's objective function was evaluated based on the number of users in the network, as shown in Figure \ref{fig:5} for $M=32$. The proposed scheme and the DDPG and PPO algorithms achieve converging rewards. The proposed scheme demonstrates superior resource optimization performance, surpassing the other two algorithms. In terms of energy efficiency, the proposed scheme attains the highest rate at 625 bits per joule per hertz. In comparison, the PPO algorithm starts at 450 bits per joule per hertz and converges to approximately 580 bits per joule per hertz after around 400 iterations.

Moreover, the DDPG algorithm achieves the same energy efficiency as the proposed scheme. Conversely, the Random Policy algorithm exhibits poor convergence, with its curve fluctuating between 400 and 500 bits per joule per hertz. Compared to the proposed scheme, the other algorithms demonstrate weaker stability and inferior performance in optimizing the objective function.

{The advantage of the proposed scheme (MADDPG) lies in its ability to achieve both high resource optimization performance and energy efficiency, as demonstrated in Figure \ref{fig:5}. The proposed scheme outperforms the benchmark algorithms, showcasing its superior stability and consistent performance. It achieves the highest energy efficiency rate of 625 bits per joule per hertz, providing significant gains compared to the PPO algorithm's convergence to around 580 bits per joule per hertz after approximately 400 iterations. Thus, the proposed scheme (MADDPG) stands out as an effective and efficient solution for optimizing system objectives in complex multi-agent networks.}
\par
The relationship between the users in a network and the objective function value of algorithms is demonstrated in Figure \ref{fig:6}. Results demonstrate that the system's energy efficiency increases as the user density per BS and the satellite increase. The proposed scheme outperforms other benchmark algorithms regarding energy efficiency when three to five users share each BS and satellite. With each BS and satellite having five users, the system's energy efficiency can reach up to 1180 bits/Joule/Hz. The proposed scheme is significantly more energy-efficient than the random policy, DDPG, PPO, and GA algorithms, indicating its effectiveness.

Figure \ref{fig:7} shows the convergence analysis of Algorithm 2 for various local capacities and file libraries. With just 1000 training iterations, Algorithm 2 converges quickly, taking around 60 iterations. The figure compares four examples with different file library sizes and local capacities.

{Similarly, the convergence analysis of Algorithm 2 is investigated as the sizes of the satellite local capacity, BS, and file library vary. Algorithm 2 performs well with four different capacities. However, ineffective local cache deployment in the first training iteration increases power consumption, yielding a decrease in cache reward to below 1090 bits per Joule per Hertz. The cache rewards improve as training iterations increase and converge after about 50 iterations. As the file library grows, the cache reward in the framework decreases due to the fixed local cache capacity, making it more challenging to locate the file in Algorithm 2. The local cache capacity and reward decline for the same cache file library.}

Figures \ref{fig:8} and \ref{fig:9} compare the convergence processes for energy efficiency and cache hit rate among various cache optimization algorithms at $N_f=3$ and $U=40$. As illustrated in Figure \ref{fig:8}, Algorithm 2 and the DDPG algorithm can converge in energy efficiency. However, Algorithm 2 performs better for a given cache and file library size than other algorithms. The DDPG algorithm attains a training value of 1100 bits per joule per Hz at around 700 iterations, while benchmark-3 exhibits an oscillating curve between 600 and 650 bits per joule per Hz, hindering convergence. Algorithm 2 outperforms other algorithms in optimizing the system's objective function more effectively and steadily.

Figure \ref{fig:9} shows the Algorithm 2 request competition rate. Results demonstrate that the value rises to 0.13 and converges at 0.33, while the DDPG algorithm's cache hit rate converges slowly. Therefore, the proposed scheme produces better cache hit rates.

Figures \ref{fig:10} and \ref{fig:11} show the request competition rate and the energy efficiency of various algorithms for each base station and satellite with different cache sizes. Despite the uncached strategy requiring less memory, the MADDPG algorithm has lower energy efficiency when the cache capacity is only 1, as depicted in Figure \ref{fig:10}. This trend is due to the limited serving capacity if the local cache acts as a performance bottleneck in retrieving the relevant file as per the user's request, resulting in suboptimal outcomes. However, as the cache capacity varies from value 1 to value 6, the performance of the proposed scheme becomes more energy-efficient. The MADDPG algorithm consistently outperforms other algorithms regarding energy efficiency, and the performance gap between the two algorithms widens with larger cache sizes.

{Figure \ref{fig:11} shows that the proposed MADDPG algorithm achieves a higher cache hit rate than other algorithms, following the same pattern as Figure \ref{fig:10}. These graphs help us better understand the dynamics of the relationships. The cache reward and hit rate will decline proportionally to the size of the file library within a specific range, assuming that the local cache capacity will not change. This is because accessing the necessary files will get harder and harder as the library gets more significant. On the other hand, increasing the local cache capacity will increase both cache reward and cache hit rate, presuming that the size of the file library stays constant.}

These findings validate the effectiveness of the proposed MADDPG scheme in achieving high energy efficiency and cache hit rates, outperforming other benchmark algorithms across various network configurations and cache sizes.
\section{Conclusion and Future Directions}
{In conclusion, our proposed approach for enhancing energy efficiency in integrated terrestrial and satellite NOMA communication networks offers several advantages compared to existing reference contributions. Firstly, our use of a multi-agent deep reinforcement learning technique, specifically the MADDPG algorithm, outperforms benchmark algorithms and the standard DDPG algorithm, which only utilizes a single agent. This highlights the effectiveness of our approach in addressing the complexities of resource allocation and cache design in a multi-agent setting. By leveraging the MADDPG algorithm, we achieve optimal user association, power management, and cache layout. It allows us to model users and BSs as agents and enables the resource management and cache design optimization solution in a more efficient way. Moreover, the proposed MADDPG-based cache design strategy enables BSs and satellites to act as agents and intelligently select files from the library to store in local cache pools. This strategy improves the efficiency of data retrieval and maximizes the energy efficiency of the network. Our results demonstrate the benefits of the considered framework of energy efficiency optimization in integrated terrestrial-satellite NOMA networks. We significantly advance this research area by outperforming the benchmark algorithms and utilizing multi-agent deep reinforcement learning. Our work serves as a solid foundation for future studies and opens up opportunities to explore more complex scenarios, such as the optimal allocation of resources in a multiple-layer NOMA-enabled satellite communication network model. Specifically, we aim to focus on minimizing power consumption, which is crucial for sustainable and energy-efficient network operations.}
\par
{While our proposed model showcases various advantages, there are certain limitations that should be acknowledged. One limitation is the computational complexity associated with multi-agent deep reinforcement learning algorithms, which can require substantial computational resources and time. Additionally, the performance of our approach heavily relies on accurate modelling and representation of the network environment, including user behaviour and channel conditions. Ensuring the reliability and real-world applicability of these models is an ongoing challenge. In summary, our proposed approach offers significant advantages over existing contributions, including superior performance, efficient multi-agent optimization, and effective cache design. While there are limitations to be addressed, our research paves the way for future investigations and advancements in energy-efficient integrated terrestrial and satellite communication networks.}
\bibliographystyle{IEEEtran}
\bibliography{ReferenceBibFile}

\begin{thebibliography}{10}
\providecommand{\url}[1]{#1}
\csname url@samestyle\endcsname
\providecommand{\newblock}{\relax}
\providecommand{\bibinfo}[2]{#2}
\providecommand{\BIBentrySTDinterwordspacing}{\spaceskip=0pt\relax}
\providecommand{\BIBentryALTinterwordstretchfactor}{4}
\providecommand{\BIBentryALTinterwordspacing}{\spaceskip=\fontdimen2\font plus
\BIBentryALTinterwordstretchfactor\fontdimen3\font minus
  \fontdimen4\font\relax}
\providecommand{\BIBforeignlanguage}[2]{{%
\expandafter\ifx\csname l@#1\endcsname\relax
\typeout{** WARNING: IEEEtran.bst: No hyphenation pattern has been}%
\typeout{** loaded for the language `#1'. Using the pattern for}%
\typeout{** the default language instead.}%
\else
\language=\csname l@#1\endcsname
\fi
#2}}
\providecommand{\BIBdecl}{\relax}
\BIBdecl

\bibitem{khan2023ris}
W.~U. Khan, E.~Lagunas, A.~Mahmood, S.~Chatzinotas, and B.~Ottersten,
  ``{RIS-assisted energy-efficient LEO satellite communications with NOMA},''
  \emph{arXiv preprint arXiv:2306.10422}, 2023.

\bibitem{9351705}
B.~Cao, M.~Li, X.~Liu, J.~Zhao, W.~Cao, and Z.~Lv, ``Many-objective deployment
  optimization for a drone-assisted camera network,'' \emph{IEEE Transactions
  on Network Science and Engineering}, vol.~8, no.~4, pp. 2756--2764, October
  2021.

\bibitem{9640472}
B.~Li, M.~Zhang, Y.~Rong, and Z.~Han, ``{Transceiver optimization for wireless
  powered time-division duplex MU-MIMO systems: Non-robust and robust
  designs},'' \emph{IEEE Transactions on Wireless Communications}, vol.~21,
  no.~6, pp. 4594--4607, June 2022.

\bibitem{geraci2022integrating}
G.~Geraci, D.~Lopez-Perez, M.~Benzaghta, and S.~Chatzinotas, ``{Integrating
  terrestrial and non-terrestrial networks: 3D opportunities and challenges},''
  \emph{IEEE Communications Magazine}, 2022.

\bibitem{9325911}
B.~Cao, Z.~Sun, J.~Zhang, and Y.~Gu, ``{Resource allocation in 5G IoV
  architecture based on SDN and gog-cloud computing},'' \emph{IEEE Transactions
  on Intelligent Transportation Systems}, vol.~22, no.~6, pp. 3832--3840, June
  2021.

\bibitem{khan2023rate}
W.~U. Khan, Z.~Ali, E.~Lagunas, A.~Mahmood, M.~Asif, A.~Ihsan, S.~Chatzinotas,
  B.~Ottersten, and O.~A. Dobre, ``Rate splitting multiple access for next
  generation cognitive radio enabled {LEO} satellite networks,'' \emph{IEEE
  Transactions on Wireless Communications}, pp. 1--1, 2023.

\bibitem{azari2022evolution}
M.~M. Azari, S.~Solanki, S.~Chatzinotas, O.~Kodheli, H.~Sallouha, A.~Colpaert,
  J.~F.~M. Montoya, S.~Pollin, A.~Haqiqatnejad, A.~Mostaani \emph{et~al.},
  ``{Evolution of non-terrestrial networks from 5G to 6G: A survey},''
  \emph{IEEE communications surveys \& tutorials}, 2022.

\bibitem{mahmood2022weighted}
A.~Mahmood, A.~Ahmed, M.~Naeem, M.~R. Amirzada, and A.~Al-Dweik, ``Weighted
  utility aware computational overhead minimization of wireless power mobile
  edge cloud,'' \emph{Computer Communications}, vol. 190, pp. 178--189, 2022.

\bibitem{saafi2022ai}
S.~Saafi, O.~Vikhrova, G.~Fodor, J.~Hosek, and S.~Andreev, ``{AI-aided
  integrated terrestrial and non-terrestrial 6G solutions for sustainable
  maritime networking},'' \emph{IEEE Network}, vol.~36, no.~3, pp. 183--190,
  2022.

\bibitem{khan2019joint}
W.~U. Khan, F.~Jameel, T.~Ristaniemi, S.~Khan, G.~A.~S. Sidhu, and J.~Liu,
  ``{Joint spectral and energy efficiency optimization for downlink NOMA
  networks},'' \emph{IEEE Transactions on Cognitive Communications and
  Networking}, vol.~6, no.~2, pp. 645--656, 2019.

\bibitem{zhang2018energy}
H.~Zhang, B.~Wang, C.~Jiang, K.~Long, A.~Nallanathan, V.~C. Leung, and H.~V.
  Poor, ``Energy efficient dynamic resource optimization in noma system,''
  \emph{IEEE Transactions on Wireless Communications}, vol.~17, no.~9, pp.
  5671--5683, 2018.

\bibitem{zhang2020energy}
H.~Zhang, H.~Zhang, W.~Liu, K.~Long, J.~Dong, and V.~C. Leung, ``Energy
  efficient user clustering, hybrid precoding and power optimization in
  terahertz mimo-noma systems,'' \emph{IEEE Journal on selected areas in
  communications}, vol.~38, no.~9, pp. 2074--2085, 2020.

\bibitem{khan2021energy}
A.~Nauman, M.~Obayya, M.~M. Asiri, K.~Yadav, M.~Maashi, M.~Assiri, M.~K. Ehsan,
  and S.~W. Kim, ``Minimizing energy consumption for noma multi-drone
  communications in automotive-industry 5.0,'' \emph{Journal of King Saud
  University-Computer and Information Sciences}, p. 101547, 2023.

\bibitem{9001132}
A.~Mahmood, A.~Ahmed, M.~Naeem, and Y.~Hong, ``Partial offloading in energy
  harvested mobile edge computing: A direct search approach,'' \emph{IEEE
  Access}, vol.~8, pp. 36\,757--36\,763, 2020.

\bibitem{zhu2017non}
X.~Zhu, C.~Jiang, L.~Kuang, N.~Ge, and J.~Lu, ``Non-orthogonal multiple access
  based integrated terrestrial-satellite networks,'' \emph{IEEE Journal on
  Selected Areas in Communications}, vol.~35, no.~10, pp. 2253--2267, 2017.

\bibitem{khan2020spectral}
W.~U. Khan, J.~Liu, F.~Jameel, V.~Sharma, R.~J{\"a}ntti, and Z.~Han,
  ``{Spectral efficiency optimization for next generation NOMA-enabled IoT
  networks},'' \emph{IEEE Transactions on Vehicular Technology}, vol.~69,
  no.~12, pp. 15\,284--15\,297, 2020.

\bibitem{fu2020integrated}
S.~Fu, J.~Gao, and L.~Zhao, ``Integrated resource management for
  terrestrial-satellite systems,'' \emph{IEEE Transactions on Vehicular
  Technology}, vol.~69, no.~3, pp. 3256--3266, 2020.

\bibitem{deng2019joint}
B.~Deng, C.~Jiang, J.~Yan, N.~Ge, S.~Guo, and S.~Zhao, ``Joint multigroup
  precoding and resource allocation in integrated terrestrial-satellite
  networks,'' \emph{IEEE Transactions on Vehicular Technology}, vol.~68, no.~8,
  pp. 8075--8090, 2019.

\bibitem{9672696}
Z.~Zhao, G.~Xu, N.~Zhang, and Q.~Zhang, ``Performance analysis of the hybrid
  satellite-terrestrial relay network with opportunistic scheduling over
  generalized fading channels,'' \emph{IEEE Transactions on Vehicular
  Technology}, vol.~71, no.~3, pp. 2914--2924, March 2022.

\bibitem{9552222}
S.~Pan, M.~Lin, M.~Xu, S.~Zhu, L.-A. Bian, and G.~Li, ``A low-profile
  programmable beam scanning holographic array antenna without phase
  shifters,'' \emph{IEEE Internet of Things Journal}, vol.~9, no.~11, pp.
  8838--8851, June 2022.

\bibitem{9652043}
B.~Li, Q.~Li, Y.~Zeng, Y.~Rong, and R.~Zhang, ``{3D trajectory optimization for
  energy-efficient UAV communication: A control design perspective},''
  \emph{IEEE Transactions on Wireless Communications}, vol.~21, no.~6, pp.
  4579--4593, June 2022.

\bibitem{giordani2020non}
M.~Giordani and M.~Zorzi, ``{Non-terrestrial networks in the 6G era: Challenges
  and opportunities},'' \emph{IEEE Network}, vol.~35, no.~2, pp. 244--251,
  2020.

\bibitem{sattarzadeh2021satellite}
A.~Sattarzadeh, Y.~Liu, A.~Mohamed, R.~Song, P.~Xiao, Z.~Song, H.~Zhang,
  R.~Tafazolli, and C.~Niu, ``{Satellite-based non-terrestrial networks in 5G:
  Insights and challenges},'' \emph{IEEE Access}, vol.~10, pp.
  11\,274--11\,283, 2021.

\bibitem{rinaldi2020non}
F.~Rinaldi, H.-L. Maattanen, J.~Torsner, S.~Pizzi, S.~Andreev, A.~Iera,
  Y.~Koucheryavy, and G.~Araniti, ``{Non-terrestrial networks in 5G \& beyond:
  A survey},'' \emph{IEEE access}, vol.~8, pp. 165\,178--165\,200, 2020.

\bibitem{cao2020deep}
Y.~Cao, S.-Y. Lien, and Y.-C. Liang, ``{Deep reinforcement learning for
  multi-user access control in non-terrestrial networks},'' \emph{IEEE
  Transactions on Communications}, vol.~69, no.~3, pp. 1605--1619, 2020.

\bibitem{khan2022opportunities}
W.~U. Khan, A.~Mahmood, A.~Bozorgchenani, M.~A. Jamshed, A.~Ranjha, E.~Lagunas,
  H.~Pervaiz, S.~Chatzinotas, B.~Ottersten, and P.~Popovski, ``Opportunities
  for intelligent reflecting surfaces in 6g-empowered v2x communications,''
  \emph{arXiv preprint arXiv:2210.00494}, 2022.

\bibitem{ahmad2022security}
I.~Ahmad, J.~Suomalainen, P.~Porambage, A.~Gurtov, J.~Huusko, and
  M.~H{\"o}yhty{\"a}, ``{Security of Satellite-Terrestrial Communications:
  Challenges and Potential Solutions},'' \emph{IEEE Access}, vol.~10, pp.
  96\,038--96\,052, 2022.

\bibitem{raza2021task}
S.~Raza, S.~Wang, M.~Ahmed, M.~R. Anwar, M.~A. Mirza, and W.~U. Khan, ``Task
  offloading and resource allocation for iov using 5g nr-v2x communication,''
  \emph{IEEE Internet of Things Journal}, vol.~9, no.~13, pp. 10\,397--10\,410,
  2021.

\bibitem{ahmed2023vehicular}
M.~Ahmed \emph{et~al.}, ``Vehicular communication network enabled cav data
  offloading: A review,'' \emph{IEEE Transactions on Intelligent Transportation
  Systems}, 2023.

\bibitem{rasheed2022lstm}
I.~Rasheed \emph{et~al.}, ``Lstm-based distributed conditional generative
  adversarial network for data-driven 5g-enabled maritime uav communications,''
  \emph{IEEE Transactions on Intelligent Transportation Systems}, 2022.

\bibitem{shome2022federated}
D.~Shome \emph{et~al.}, ``Federated learning and next generation wireless
  communications: A survey on bidirectional relationship,'' \emph{Transactions
  on Emerging Telecommunications Technologies}, vol.~33, no.~7, p. e4458, 2022.

\bibitem{mahmood2023joint}
A.~Mahmood, T.~X. Vu, S.~Chatzinotas, and B.~Ottersten, ``Joint optimization of
  3d placement and radio resource allocation for per-uav sum rate
  maximization,'' \emph{IEEE Transactions on Vehicular Technology}, 2023.

\bibitem{hasan2022securing}
T.~Hasan \emph{et~al.}, ``Securing industrial internet of things against botnet
  attacks using hybrid deep learning approach,'' \emph{IEEE Transactions on
  Network Science and Engineering}, 2022.

\bibitem{jiao2020network}
J.~Jiao, Y.~Sun, S.~Wu, Y.~Wang, and Q.~Zhang, ``{Network utility maximization
  resource allocation for NOMA in satellite-based Internet of Things},''
  \emph{IEEE Internet of Things Journal}, vol.~7, no.~4, pp. 3230--3242, 2020.

\bibitem{wang2020noma}
A.~Wang, L.~Lei, E.~Lagunas, A.~I. P{\'e}rez-Neira, S.~Chatzinotas, and
  B.~Ottersten, ``{NOMA-enabled multi-beam satellite systems: Joint
  optimization to overcome offered-requested data mismatches},'' \emph{IEEE
  Transactions on Vehicular Technology}, vol.~70, no.~1, pp. 900--913, 2020.

\bibitem{ge2021joint}
R.~Ge, D.~Bian, J.~Cheng, K.~An, J.~Hu, and G.~Li, ``{Joint user pairing and
  power allocation for noma-based geo and leo satellite network},'' \emph{IEEE
  Access}, vol.~9, pp. 93\,255--93\,266, 2021.

\bibitem{wang2020admission}
R.~Wang, W.~Kang, G.~Liu, R.~Ma, and B.~Li, ``{Admission control and power
  allocation for NOMA-based satellite multi-beam network},'' \emph{IEEE
  Access}, vol.~8, pp. 33\,631--33\,643, 2020.

\bibitem{ji2020popularity}
Z.~Ji, S.~Wu, C.~Jiang, and W.~Wang, ``Popularity-driven content placement and
  multi-hop delivery for terrestrial-satellite networks,'' \emph{IEEE
  Communications Letters}, vol.~24, no.~11, pp. 2574--2578, 2020.

\bibitem{lagunas2019power}
E.~Lagunas, L.~Lei, S.~Chatzinotas, and B.~Ottersten, ``Power and flow
  assignment for 5g integrated terrestrial-satellite backhaul networks,'' in
  \emph{2019 IEEE Wireless Communications and Networking Conference
  (WCNC)}.\hskip 1em plus 0.5em minus 0.4em\relax IEEE, 2019, pp. 1--6.

\bibitem{shaat2017joint}
M.~Shaat, A.~I. P{\'e}rez-Neira, G.~Femenias, and F.~Riera-Palou, ``Joint
  frequency assignment and flow control for hybrid terrestrial-satellite
  backhauling networks,'' in \emph{2017 International Symposium on Wireless
  Communication Systems (ISWCS)}.\hskip 1em plus 0.5em minus 0.4em\relax IEEE,
  2017, pp. 293--298.

\bibitem{gao2021sum}
Z.~Gao, A.~Liu, C.~Han, and X.~Liang, ``Sum rate maximization of massive mimo
  noma in leo satellite communication system,'' \emph{IEEE Wireless
  Communications Letters}, vol.~10, no.~8, pp. 1667--1671, 2021.

\bibitem{liao2020distributed}
X.~Liao, X.~Hu, Z.~Liu, S.~Ma, L.~Xu, X.~Li, W.~Wang, and F.~M. Ghannouchi,
  ``Distributed intelligence: A verification for multi-agent drl-based
  multibeam satellite resource allocation,'' \emph{IEEE Communications
  Letters}, vol.~24, no.~12, pp. 2785--2789, 2020.

\bibitem{hu2018deep}
X.~Hu, S.~Liu, R.~Chen, W.~Wang, and C.~Wang, ``A deep reinforcement
  learning-based framework for dynamic resource allocation in multibeam
  satellite systems,'' \emph{IEEE Communications Letters}, vol.~22, no.~8, pp.
  1612--1615, 2018.

\bibitem{khan2021backscatter}
W.~U. Khan, F.~Jameel, N.~Kumar, R.~J{\"a}ntti, and M.~Guizani,
  ``{Backscatter-enabled efficient V2X communication with non-orthogonal
  multiple access},'' \emph{IEEE Transactions on Vehicular Technology},
  vol.~70, no.~2, pp. 1724--1735, 2021.

\bibitem{khan2021backscatterL}
W.~U. Khan, X.~Li, M.~Zeng, and O.~A. Dobre, ``{Backscatter-enabled NOMA for
  future 6G systems: A new optimization framework under imperfect SIC},''
  \emph{IEEE Communications Letters}, vol.~25, no.~5, pp. 1669--1672, 2021.

\bibitem{khan2021joint}
W.~U. Khan, F.~Jameel, X.~Li, M.~Bilal, and T.~A. Tsiftsis, ``{Joint spectrum
  and energy optimization of NOMA-enabled small-cell networks with QoS
  guarantee},'' \emph{IEEE Transactions on Vehicular Technology}, vol.~70,
  no.~8, pp. 8337--8342, 2021.

\bibitem{khan2022nomaInd5}
W.~U. Khan, A.~Ihsan, T.~N. Nguyen, Z.~Ali, and M.~A. Javed, ``{NOMA-enabled
  backscatter communications for green transportation in automotive-industry
  5.0},'' \emph{IEEE Transactions on Industrial Informatics}, vol.~18, no.~11,
  pp. 7862--7874, 2022.

\bibitem{khan2023integration}
W.~U. Khan, E.~Lagunas, A.~Mahmood, Z.~Ali, M.~Asif, S.~Chatzinotas, and
  B.~Ottersten, ``Integration of noma with reflecting intelligent surfaces: A
  multi-cell optimization with sic decoding errors,'' \emph{IEEE Transactions
  on Green Communications and Networking}, 2023.

\bibitem{khan2022integration}
W.~U. Khan, E.~Lagunas, A.~Mahmood, Z.~Ali, S.~Chatzinotas, B.~Ottersten, and
  O.~A. Dobre, ``Integration of backscatter communication with multi-cell noma:
  a spectral efficiency optimization under imperfect sic,'' in \emph{2022 IEEE
  27th International Workshop on Computer Aided Modeling and Design of
  Communication Links and Networks (CAMAD)}.\hskip 1em plus 0.5em minus
  0.4em\relax IEEE, 2022, pp. 147--152.

\bibitem{zhang2020power}
H.~Zhang, N.~Yang, W.~Huangfu, K.~Long, and V.~C. Leung, ``Power control based
  on deep reinforcement learning for spectrum sharing,'' \emph{IEEE
  Transactions on Wireless Communications}, vol.~19, no.~6, pp. 4209--4219,
  2020.

\bibitem{geraci2022will}
G.~Geraci, A.~Garcia-Rodriguez, M.~M. Azari, A.~Lozano, M.~Mezzavilla,
  S.~Chatzinotas, Y.~Chen, S.~Rangan, and M.~Di~Renzo, ``What will the future
  of uav cellular communications be? a flight from 5g to 6g,'' \emph{IEEE
  communications surveys \& tutorials}, vol.~24, no.~3, pp. 1304--1335, 2022.

\bibitem{lin20215g}
X.~Lin, S.~Rommer, S.~Euler, E.~A. Yavuz, and R.~S. Karlsson, ``5g from space:
  An overview of 3gpp non-terrestrial networks,'' \emph{IEEE Communications
  Standards Magazine}, vol.~5, no.~4, pp. 147--153, 2021.

\bibitem{lin2022overview}
X.~Lin, ``An overview of 5g advanced evolution in 3gpp release 18,'' \emph{IEEE
  Communications Standards Magazine}, vol.~6, no.~3, pp. 77--83, 2022.

\bibitem{ferreirment}
P.~V.~R. Ferreira, R.~Paffenroth, A.~M. Wyglinski, T.~M. Hackett, S.~G. Bilen,
  R.~C. Reinhart, and D.~J. Mortensen, ``Reinforcement learning for satellite
  communications: From leo to deep space operations,'' \emph{IEEE
  Communications Magazine}, vol.~57, no.~5, pp. 70--75, 2019.

\bibitem{zhong2020deep}
C.~Zhong, M.~C. Gursoy, and S.~Velipasalar, ``Deep reinforcement learning-based
  edge caching in wireless networks,'' \emph{IEEE Transactions on Cognitive
  Communications and Networking}, vol.~6, no.~1, pp. 48--61, 2020.

\bibitem{zhang2019double}
Z.~Zhang, H.~Chen, M.~Hua, C.~Li, Y.~Huang, and L.~Yang, ``Double coded caching
  in ultra dense networks: Caching and multicast scheduling via deep
  reinforcement learning,'' \emph{IEEE Transactions on Communications},
  vol.~68, no.~2, pp. 1071--1086, 2019.

\bibitem{qian202ment}
Y.~Qian, R.~Wang, J.~Wu, B.~Tan, and H.~Ren, ``Reinforcement learning-based
  optimal computing and caching in mobile edge network,'' \emph{IEEE Journal on
  Selected Areas in Communications}, vol.~38, no.~10, pp. 2343--2355, 2020.

\bibitem{zhang2021joint}
T.~Zhang, Z.~Wang, Y.~Liu, W.~Xu, and A.~Nallanathan, ``Joint resource,
  deployment, and caching optimization for ar applications in dynamic uav noma
  networks,'' \emph{IEEE Transactions on Wireless Communications}, vol.~21,
  no.~5, pp. 3409--3422, 2021.

\bibitem{zhang20caching}
------, ``Caching placement and resource allocation for cache-enabling uav noma
  networks,'' \emph{IEEE Transactions on Vehicular Technology}, vol.~69,
  no.~11, pp. 12\,897--12\,911, 2020.

\bibitem{li2021multi}
X.~Li, H.~Zhang, W.~Li, and K.~Long, ``Multi-agent drl for user association and
  power control in terrestrial-satellite network,'' in \emph{2021 IEEE Global
  Communications Conference (GLOBECOM)}.\hskip 1em plus 0.5em minus 0.4em\relax
  IEEE, 2021, pp. 1--5.

\bibitem{9992172}
G.~Geraci, D.~López-Pérez, M.~Benzaghta, and S.~Chatzinotas, ``Integrating
  terrestrial and non-terrestrial networks: 3d opportunities and challenges,''
  \emph{IEEE Communications Magazine}, vol.~61, no.~4, pp. 42--48, 2023.

\bibitem{khan2020secureMC}
W.~U. Khan, J.~Liu, F.~Jameel, M.~T.~R. Khan, S.~H. Ahmed, and R.~J{\"a}ntti,
  ``{Secure backscatter communications in multi-cell NOMA networks: Enabling
  link security for massive IoT networks},'' in \emph{IEEE INFOCOM 2020-IEEE
  Conference on Computer Communications Workshops (INFOCOM WKSHPS)}.\hskip 1em
  plus 0.5em minus 0.4em\relax IEEE, 2020, pp. 213--218.

\bibitem{9712216}
H.~Jiang, X.~Dai, Z.~Xiao, and A.~Iyengar, ``Joint task offloading and resource
  allocation for energy-constrained mobile edge computing,'' \emph{IEEE
  Transactions on Mobile Computing}, vol.~22, no.~7, pp. 4000--4015, July 2023.

\bibitem{khan2021noma}
W.~U. Khan, X.~Li, A.~Ihsan, M.~A. Khan, V.~G. Menon, and M.~Ahmed,
  ``{NOMA-enabled optimization framework for next-generation small-cell IoV
  networks under imperfect SIC decoding},'' \emph{IEEE Transactions on
  Intelligent Transportation Systems}, vol.~23, no.~11, pp. 22\,442--22\,451,
  2021.

\bibitem{liu2021uav}
Q.~Liu, R.~Liu, Z.~Wang, and J.~S. Thompson, ``Uav swarm-enabled localization
  in isolated region: a rigidity-constrained deployment perspective,''
  \emph{IEEE Wireless Communications Letters}, vol.~10, no.~9, pp. 2032--2036,
  2021.

\bibitem{liu2021v2x}
Q.~Liu, R.~Liu, Z.~Wang, L.~Han, and J.~S. Thompson, ``A v2x-integrated
  positioning methodology in ultradense networks,'' \emph{IEEE Internet of
  Things Journal}, vol.~8, no.~23, pp. 17\,014--17\,028, 2021.

\bibitem{liu2023management}
Q.~Liu, R.~Liu, Y.~Zhang, Y.~Yuan, Z.~Wang, H.~Yang, L.~Ye, M.~Guizani, and
  J.~S. Thompson, ``Management of positioning functions in cellular networks
  for time-sensitive transportation applications,'' \emph{IEEE Transactions on
  Intelligent Transportation Systems}, 2023.

\bibitem{9375493}
B.~Cao, J.~Zhang, X.~Liu, Z.~Sun, W.~Cao, R.~M. Nowak, and Z.~Lv,
  ``Edge–cloud resource scheduling in space–air–ground-integrated
  networks for internet of vehicles,'' \emph{IEEE Internet of Things Journal},
  vol.~9, no.~8, pp. 5765--5772, April 2022.

\bibitem{khan2022rateconf}
W.~U. Khan, Z.~Ali, E.~Lagunas, S.~Chatzinotas, and B.~Ottersten, ``{Rate
  Splitting Multiple Access for Cognitive Radio GEO-LEO Co-Existing Satellite
  Networks},'' in \emph{GLOBECOM 2022-2022 IEEE Global Communications
  Conference}.\hskip 1em plus 0.5em minus 0.4em\relax IEEE, 2022, pp.
  5165--5170.

\bibitem{arulku7deep}
K.~Arulkumaran, M.~P. Deisenroth, M.~Brundage, and A.~A. Bharath, ``Deep
  reinforcement learning: A brief survey,'' \emph{IEEE Signal Processing
  Magazine}, vol.~34, no.~6, pp. 26--38, 2017.

\end{thebibliography}
\end{document}